\documentclass[aps,prb,floatfix,twocolumn,showpacs,nofitinbib,hyperref=pdftex,citeautoscript]{revtex4-1}
\usepackage{epsfig}
\usepackage{amsmath}
\usepackage{amssymb}
\usepackage{amsfonts}
\usepackage{ dsfont }
\usepackage{ comment,color }
\usepackage{lipsum}
\usepackage{hyperref}

\hypersetup{colorlinks=true,linkcolor=blue,anchorcolor=blue,citecolor=blue,filecolor=blue,urlcolor=blue,bookmarksnumbered=true,pdfview=FitB}

\newcommand{\beqa}{\begin{eqnarray}}
\newcommand{\eeqa}{\end{eqnarray}}

\begin{document}

\hsize\textwidth\columnwidth\hsize\csname@twocolumnfalse\endcsname

\title{Low-field Topological Threshold in Majorana Double Nanowires}
\author{Constantin Schrade, Manisha Thakurathi, Christopher Reeg, Silas Hoffman,  Jelena Klinovaja, and Daniel Loss}

\affiliation{Department of Physics, University of Basel,
Klingelbergstrasse 82, CH-4056 Basel, Switzerland}

\date{\today}

\vskip1.5truecm
\begin{abstract}
A hard proximity-induced superconducting gap has recently been observed in semiconductor nanowire systems at low magnetic fields. However, in the topological regime at high magnetic fields, a soft gap emerges and represents a fundamental obstacle to topologically protected quantum information processing with Majorana bound states. Here we show that in a setup of double Rashba nanowires that are coupled to an $s$-wave superconductor and subjected to an external magnetic field along the wires, the topological threshold can be significantly reduced by the destructive interference of direct and crossed-Andreev pairing in this setup, precisely down to the magnetic field regime in which current experimental technology allows for a hard superconducting gap. We also show that the resulting Majorana bound states exhibit sufficiently short localization lengths, which makes them ideal candidates for future braiding experiments.
\end{abstract}


\maketitle
\section{Introduction}
Majorana bound states (MBSs) form the building blocks of a topologically protected qubit. Over the last years, the first generation of Majorana devices were fabricated based on an $s$-wave superconductor (SC) proximity-coupled either to a nanowire (NW) with Rashba spin-orbit interaction (SOI) in the presence of a strong magnetic field \cite{bib:Lutchyn2010,bib:Oreg2010,bib:Mourik2012,bib:Das2012,bib:Rokhinson2012,bib:Deng2013} or to a chain of magnetic atoms \cite{bib:Klinovaja2013,bib:Braunecker2013,bib:Pientka2013,bib:Nadj-Perge2013,bib:Nadj-Perge2014,bib:Ruby2015,bib:Pawlak2016}. 
These devices provided the first experimental signatures of MBSs in the form of zero-bias conductance peaks~\cite{bib:Mourik2012,bib:Das2012,bib:Rokhinson2012,bib:Deng2013,bib:Nadj-Perge2014,bib:Ruby2015,bib:Pawlak2016}.
Today, the most important open challenge is to perform manipulations on the MBSs which should ultimately allow for the confirmation of  their non-Abelian braiding statistics. For this  purpose, NW devices appear particularly promising, as they provide a simple means of moving MBSs by the use of local gates \cite{bib:Alicea2011}. Unfortunately, despite the plethora of experimental breakthroughs, a long-standing \cite{bib:Takei2013,bib:Stanescu2014} and still unresolved \cite{bib:Stanescu2017,bib:Reeg2017_3} obstacle to NW-based braiding experiments is that the proximity-induced superconducting gap in the NW is well-defined only for weak magnetic fields in the trivial regime (``hard gap"). For strong magnetic fields in the topological regime, a finite subgap conductance emerges (``soft gap") which destroys the topological protection
 \cite{bib:Chang2014,bib:Zhang2016,bib:Gul2017,bib:Deng2016,bib:Chen2016}. 
\begin{figure}[!t] \centering
\includegraphics[width=1\linewidth] {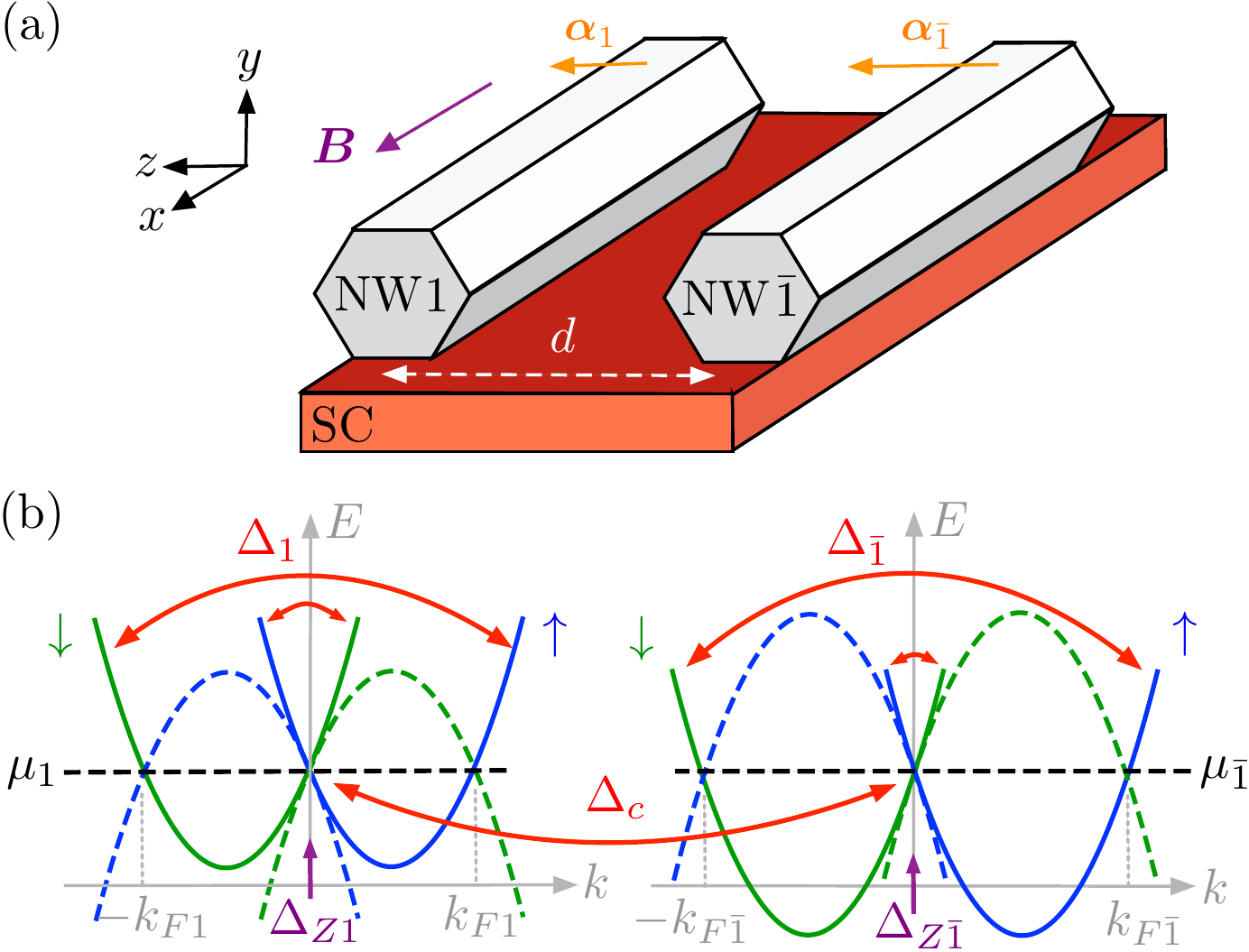}
\caption{(Color online)
(a) Two Rashba NWs (grey) labeled by $\tau=1,\bar{1}$ are aligned along the $x$ direction and proximity-coupled to an $s$-wave SC (red). Their separation in the $z$-direction is given by $d$. Both NWs are subjected to a magnetic field $\boldsymbol{B}$ which points along the $x$-axis. 
The Rashba SOI field $\boldsymbol{\alpha}_{\tau}$ in the $\tau$-wire points along the $z$-axis. 
(b) Energy spectrum in the limit of strong SOI, $E_{so,\tau}\gg\Delta_{Z\tau},\Delta_{\tau},\Delta_{c}$, with
solid (dashed) lines corresponding to electron (hole) bands. The chemical potential $\mu_{\tau}$ is tuned to the crossing point of spin-up (blue) and spin-down (green) bands in both NWs. The proximity-induced superconductivity generates a coupling between states with opposite momenta and spins belonging to the same NW (with strength $\Delta_{\tau}$) or to different NWs (with strength $\Delta_{c}$). For $|E_{so,1}-E_{so,\bar{1}}|\gg\Delta_{c}$, the 
crossed-Andreev pairing potential $\Delta_{c}$ couples only the interior branches of the spectrum at $k=0$. 
Also, the magnetic field couples states of opposite spins at $k=0$ in each NW (with strength $\Delta_{Z\tau}$). 
}\label{fig:1}
\end{figure}

Here we show that in a double NW setup the topological threshold can be reduced to the low magnetic field regime in which current experimental technology allows for a hard superconducting gap. More concretely, we consider two parallel Rashba NWs that are proximity-coupled to an $s$-wave SC and subjected to a magnetic field along the NWs, see Fig.~\ref{fig:1}(a). The SC induces both direct and crossed-Andreev pairing. We demonstrate that this double NW setup exhibits a new, previously overlooked Majorana phase that emerges at low magnetic fields. Specifically, for any finite crossed-Andreev pairing strength, we show that the system can host a single MBS at each end even when the Zeeman splitting is smaller than the strength of induced direct pairing. Notably, this phase can be realized if the direct pairing strength exceeds that of crossed-Andreev pairing, which is always the case in the absence of strong electron-electron interactions \cite{bib:Recher2002,bib:Bena2002,bib:Sato2012,bib:Klinovaja2014,bib:Klinovaja2014_2,bib:Haim2016,bib:Haim2016_2,bib:Reeg2017}.
In the limit when direct and crossed-Andreev pairing strengths are equal, we find that even an infinitesimal magnetic field can drive the system into the proposed topological phase. Interestingly, we also find that these MBSs 
exhibit a sufficiently short localization length, making them ideal for experiments on quantum information processing. 
Our theoretical proposal can readily be realized and scaled to a larger qubit architectures\cite{bib:Bravyi2010,bib:Vijay2015,bib:Vijay2016,bib:Landau2016,bib:Plugge2016,bib:Hoffman2016,bib:Vijay2016_2,bib:Plugge2017,bib:Karzig2016} in InSb/Al NW networks or can alternatively be fabricated lithographically in two-dimensional InAs/Al heterostructures \cite{bib:Suominen2017}. Consequently, it may be foundational for future experiments aimed at a controlled manipulation of MBSs.

\section{Model}
We consider a system of two parallel Rashba NWs labeled by $\tau=1,\bar{1}$, which are positioned along the $x$ direction and coupled to one another via an $s$-wave SC, see Fig.~\ref{fig:1}(a). 
The kinetic part of the Hamiltonian is given by 
\begin{align}
H_{0}=\sum_{\tau,\sigma}\int \mathrm{d}x\ \Psi_{\tau\sigma}^\dagger \left(
-\frac{\hbar^2\partial_x^2}{2m}-\mu_{\tau}\right)\Psi_{\tau\sigma'}.
\end{align}
Here, $\Psi_{\tau\sigma}(x)$ denotes the annihilation operator of an electron with mass $m$ and spin $\sigma/2=\pm1/2$ at position $x$ in the $\tau$-wire
and $\mu_{\tau}$ is the chemical potential in the $\tau$-wire.
The Rashba SOI field $\boldsymbol{\alpha}_{\tau}=\alpha_{\tau}\hat{z}$ in the $\tau$-wire is of strength $\alpha_{\tau}$ and points along the $z$ direction,
\begin{align}
H_{so}=i\sum_{\tau,\sigma,\sigma'}\alpha_{\tau}\int \mathrm{d}x\ \Psi_{\tau\sigma}^\dagger \left(\sigma_z\right)_{\sigma\sigma'}\partial_x\Psi_{\tau\sigma'},
\end{align}
where $\sigma_{x,y,z}$ are Pauli matrices acting in spin space. We assume that $\alpha_{\bar{1}}\geq\alpha_{1}>0$.
The chemical potentials in both NWs are tuned to the crossing point of the spin-polarized bands,  $\mu_{\tau}=0$. (We will address the important case when $\mu_{\tau}\neq 0$ below.) The electron bulk spectrum of $H_{0}+H_{so}$ is given by $E_{\tau\sigma}(k)=\hbar^{2}(k-\sigma k_{so,\tau})^{2}/2m-E_{so,\tau}$, where $k_{so,\tau}=m\alpha_{\tau}/\hbar^{2}$ is  the SOI wavevector
and $E_{so,\tau}=\hbar^{2}k^{2}_{so,\tau}/2m$  the SOI energy in the $\tau$-wire, see Fig.~\ref{fig:1}(b). 
Applying an external magnetic field $\boldsymbol{B}=B\hat{x}$ of magnitude $B$ parallel to the NWs induces a Zeeman splitting described by
\begin{align}
H_Z =\sum_{\tau,\sigma,\sigma'}\Delta_{Z\tau}\int \mathrm{d}x\ \Psi_{\tau\sigma}^\dagger (\sigma_x)_{\sigma\sigma'}\Psi_{\tau\sigma'},
\end{align}
where $\Delta_{Z\tau}=g_{\tau}\mu_BB/2$ is the Zeeman splitting in the $\tau$-wire, with $g_{\tau}$ the corresponding $g$-factor and $\mu_{B}$ the Bohr magneton. Assuming that the NWs are effectively one-dimensional, orbital magnetic field effects are neglected.

Superconductivity is induced in the NWs through a tunnel coupling with an $s$-wave SC. 
The tunneling of both electrons of a Cooper pair into the same NW is described by 
\begin{align}
H_{d}= \sum_{\tau,\sigma,\sigma'}
\frac{\Delta_\tau}{2} \int \mathrm{d}x \left[\Psi_{\tau\sigma}(i\sigma_{y})_{\sigma\sigma'}\Psi_{\tau\sigma'}+\text{H.c.}\right],
\end{align}
where $\Delta_\tau>0$ is the pairing potential of the induced direct superconductivity in NW $\tau$. 
Additionally, we allow for crossed-Andreev pairing, where a Cooper pair splits and one electron tunnels into each NW; this process is described by 
\begin{align}
H_{c}=\frac{\Delta_c}{2}\sum_{\tau,\sigma,\sigma'} \int \mathrm{d}x \left[\Psi_{\tau\sigma}(i\sigma_{y})_{\sigma\sigma'}\Psi_{\bar\tau\sigma'}+\text{H.c.}\right],
\label{Eq5}
\end{align}
where $\Delta_{c}>0$ is the induced crossed-Andreev pairing potential. 
The total Hamiltonian is given by $H=H_{0}+H_{so}+H_{Z}+H_{d}+H_{c}$. 
In Appendix~\ref{AppA} we provide microscopic expressions for $\Delta_{\tau}$ and $\Delta_{c}$ for the special case of weak tunnel coupling between the NWs and the SC, $\gamma\ll\Delta_{sc}$, where $\gamma$ is the energy scale of the NW-SC tunnel coupling and $\Delta_{sc}$ is the superconducting order parameter of the $s$-wave SC.
 There, we show that the ratio $\sqrt{\Delta_1\Delta_{\bar1}}/\Delta_c$ can be tuned by varying the NW separation $d$ but always satisfies $\sqrt{\Delta_1\Delta_{\bar1}}/\Delta_c >1$ in the absence of strong electron-electron interactions \cite{bib:Reeg2017}. For our discussions in the main part, we focus on the experimentally most relevant regime, $|E_{so,1}-E_{so,\bar{1}}|\gg\Delta_{Z\tau},\Delta_{\tau},\Delta_{c}\gg|\Delta_{1}-\Delta_{\bar{1}}|,|\Delta_{Z1}-\Delta_{Z\bar{1}}|$, corresponding to the limit of  strong and different SOI energies,
 with the differences in the proximity gaps and Zeeman energies being the smallest energy scales in the system. This allows us to replace $\Delta_{\tau},\Delta_{Z\tau}\rightarrow\Delta_{d},\Delta_{Z}$, and to compensate the effects of interwire tunneling, 
$H_{\Gamma}=-\Gamma\sum_{\tau,\sigma} \int \mathrm{d}x [\Psi^\dagger_{\tau\sigma}(x)\Psi_{\bar \tau\sigma}(x)+\text{H.c.}]$ with tunneling strength $\Gamma>0$, by tuning the NW chemical potentials to an appropriate {\it sweet spot}, see the stability analysis below. Notably, interwire tunneling can be substantial compared to the strength of crossed-Andreev pairing, $|\Delta_{c}/\Gamma|=\tanh(d/\xi_{sc})|\cot(k_{F,sc}d)|$ with $\xi_{sc}$ the coherence length and $k_{F,sc}$ the Fermi momentum of the $s$-wave SC,
see Appendix~\ref{AppA}. Without appropriately tuning the chemical potentials, interwire tunneling pushes the topological threshold to significantly higher magnetic fields, and not much is gained. 
For that reason, the low-field topological threshold did not emerge in previous studies \cite{bib:Gaidamauskas2014}.

\section{Topological phase diagram} 
First, we resolve the topological phase diagram of our model. We note that for $\Delta_{Z}>0$ ($\Delta_{Z}=0$) the Hamiltonian $H$ is placed in symmetry class BDI (DIII) with a $\mathbb{Z}$ ($\mathbb{Z}_{2}$) topological invariant \cite{bib:Ryu2010}.
We begin by linearizing the Hamiltonian $H_{0}+H_{so}$ around its Fermi points at $k=0$ and $k=\pm2k_{so,\tau}$ and consider the effects of magnetic field and superconductivity perturbatively, see Appendix~\ref{AppB}.  
When $|E_{so,1}-E_{so,\bar{1}}|\gg\Delta_{c}$,  the crossed-Andreev pairing couples only the interior branches, see Fig.~\ref{fig:1}(b).
We find that the spectrum is gapless at $k=0$ provided
\begin{equation}
\Delta^{2}_{c} =(\Delta_{d}\pm\Delta_{Z})^{2}.
\label{Eq6}
\end{equation}
There is no gap closing at finite-momentum for $|E_{so,1}-E_{so,\bar{1}}|\gg\Delta_{Z},\Delta_{d},\Delta_{c}$.

\begin{figure}[!t] \centering
\includegraphics[width=\linewidth] {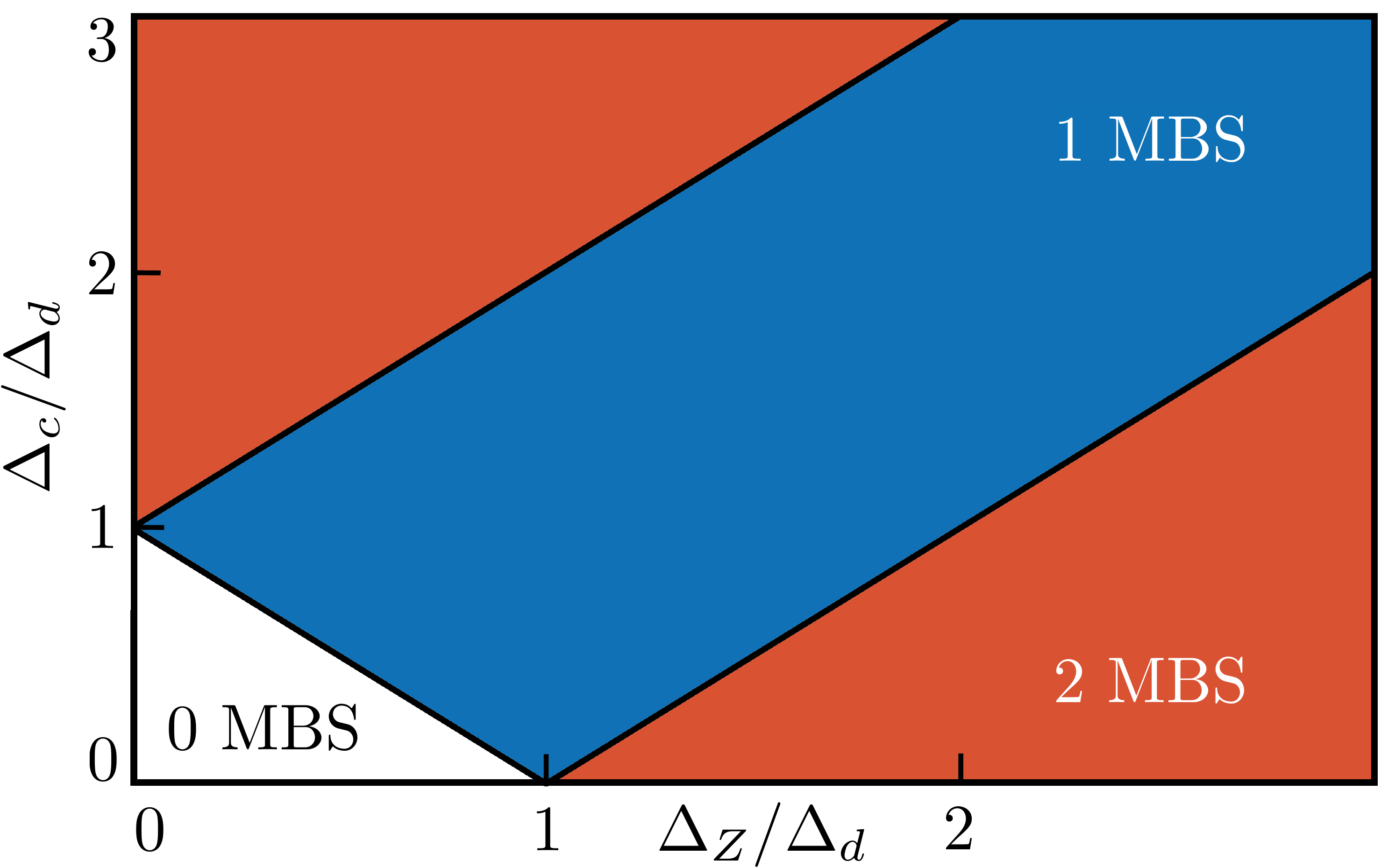}
\caption{(Color online)
Topological phase diagram as a function of the Zeeman splitting  $\Delta_{Z}$
 and the strength of the induced crossed-Andreev pairing $\Delta_{c}$ for the regime 
$ |E_{so,1}-E_{so,\bar{1}}|\gg\Delta_{Z\tau},\Delta_{\tau},\Delta_{c}\gg|\Delta_{1}-\Delta_{\bar{1}}|,|\Delta_{Z1}-\Delta_{Z\bar{1}}|$. 
There are two topological phases hosting one MBS (blue) and two MBSs (red) at each end. The trivial phase (white) does not host any MBSs. To take advantage of the low-field topological threshold, the setup shown in Fig.~\ref{fig:1}(a) should be operated in the one-MBS phase for $\Delta_{d}+\Delta_{Z}>\Delta_{c}>|\Delta_{d}-\Delta_{Z}|$. 
}\label{fig:2}
\end{figure}

Based on Eq.~\eqref{Eq6}, we are now in the position to determine the topological phases themselves, see Fig.~\ref{fig:2}. When $\Delta_{Z}=0$ and $\Delta_{c}>\Delta_{d}$ the system is a time-reversal symmetric topological superconductor and hosts a Kramers pair of MBSs at each end \cite{bib:Klinovaja2014}. For $\Delta_{Z}=0$ and all remaining values of $\Delta_{c}$ it is a trivial superconductor. Since the number of MBSs is a topological invariant, it cannot change without closing the energy gap. Consequently, for $\Delta_{c} >\Delta_{d}+\Delta_{Z}$ the system must be in a topological phase with two MBSs at each end, while for $\Delta_{d}-\Delta_{Z}>\Delta_{c}$ it must be in a trivial phase. Moreover, for $\Delta_{c}=0$ and $\Delta_{Z}>\Delta_{d}$ each NW independently hosts a pair of MBSs at its ends \cite{bib:Lutchyn2010,bib:Oreg2010}. Thus, we conclude that the system must exhibit a topological phase with two MBSs at each end for $\Delta_{Z}-\Delta_{d}>\Delta_{c}$. Finally, from an explicit calculation of the MBS wavefunctions, we find that the system hosts one MBS on each end for $\Delta_{d}+\Delta_{Z}>\Delta_{c}>|\Delta_{d}-\Delta_{Z}|$. 

We now discuss this one-MBS phase in more detail. There are three remarkable aspects:
 (1) For any finite crossed-Andreev pairing strength $\Delta_{c}$, the one-MBS phase occurs even for Zeeman splittings smaller than the direct pairing strength, $\Delta_{Z}<\Delta_{d}$. Notably, for $\Delta_{c}=\Delta_{d}$ an infinitesimal magnetic field can drive the system into the one-MBS phase. This behavior originates from the destructive interference of direct and crossed-Andreev pairing, which lowers the topological threshold to $\Delta_{Z}=\Delta_{d}-\Delta_{c}$. (2) The one-MBS phase is realized for $\Delta_c<\Delta_{d}$. This means that it can be achieved in a noninteracting system which, consequently, constitutes a powerful advantage over systems at strictly zero field which host Kramers pairs of MBSs. The latter usually rely on the presence of strong electron-electron interactions that are difficult to control experimentally \cite{bib:Klinovaja2014,bib:Klinovaja2014_2,bib:Haim2016,bib:Haim2016_2}. Also, the definition of a topological qubit in time-reversal invariant topological superconductors is potentially problematic as it requires additional symmetry conditions \cite{bib:Wolms2014,bib:Wolms2016}. Compared to that, the 
one-MBS phase allows for the standard definition of a topological qubit for topological superconductors without time-reversal symmetry. (3) The one-MBS phase exists in the limit of strong SOI, which ensures sufficiently short localization length and immediate accessibility with current experimental technology. The weak SOI limit \cite{bib:Gaidamauskas2014} is experimentally less feasible, as it leads to large localization lengths of the MBSs requiring ultra-long NWs.

\section{Localization lengths} 
\label{section4}
We continue the discussion of the one-MBS phase by computing the localization lengths of the MBS wavefunctions. 
We assume that the NW length is much longer than the MBS localization lengths, so that MBSs at opposite ends do not overlap \cite{bib:Prada2012,bib:Rainis2013,bib:Zyuzin2013}. 
The MBS wavefunctions then correspond to zero energy eigenstates of the Hamiltonian $H$ and can be determined independently for each end, see Appendix~\ref{AppC}.

We find that the MBS wavefunctions are characterized by the localization lengths determined by the two branches of the spectrum \cite{bib:Klinovaja2012}. The localization lengths corresponding to the exterior ($e$) branches at $k=\pm k_{F\tau}=\pm2k_{so,\tau}$ of the spectrum are given by the superconducting coherence lengths, $\xi_{e\tau}=\hbar\upsilon_{F\tau}/\Delta_{d}$, where $\upsilon_{F\tau}=\hbar k_{so,\tau}/m$ is the Fermi velocity in NW $\tau$. 
The localization length due to the interior ($i$) branch of the spectrum is given by 
\begin{align}
\xi_{i} &= 2\hbar\upsilon_{F1}\upsilon_{F\bar{1}}\Big[
\left(\upsilon_{F1}+\upsilon_{F\bar{1}}\right) (\Delta_{Z}-\Delta_{d})
\\
&\quad\left.+
\sqrt{
\big[(\upsilon_{F1}-\upsilon_{F\bar{1}})(\Delta_{Z}-\Delta_{d})\big]^2
+
4\upsilon_{F1}\upsilon_{F\bar{1}}  \Delta_{c}^2
}
\right]^{-1}.\nonumber
\end{align}
The total localization length is given by $\xi=\text{max}\{\xi_{i},\xi_{e\tau}\}$.

We now want to compare the MBS localization length in the double NW setup to the one
in the 
standard setup of a single Rashba NW coupled to an $s$-wave SC and subjected to a magnetic field along the NW axis \cite{bib:Lutchyn2010,bib:Oreg2010}. Assuming that the NW chemical potential is tuned to the crossing  point of the spin-polarized bands of the Rashba spectrum, this single NW setup hosts a MBS at each end provided $\Delta_{Z}>\Delta_{d}$. The MBS localization length is $\xi'=\text{max}\{\hbar\upsilon_{F}/(\Delta_{Z}-\Delta_{d}),\hbar\upsilon_{F}/\Delta_{d}\}$, where $\upsilon_{F}$ is the Fermi velocity in the NW \cite{bib:Klinovaja2012}. In general, we find that the MBS localization length of the double NW setup is always shorter than that
in the single NW setup for a fixed Zeeman splitting, $\xi\leq\xi'$ when $\Delta_{c}
>0$. To give numerical estimates, we choose typical experimental values for semiconducting NWs, $\Delta_{d}=0.1$~meV, 
$g=2$, and 
$\upsilon_{F1}=\upsilon_{F}=1.5\times10^{4}$~m/s and $ \upsilon_{F\bar{1}}=2.5\times10^{4}$~m/s.
Furthermore, we take $\Delta_{Z}=0.13$~meV for the Zeeman splitting which corresponds to a field strength of $\sim2.2$~T. For the MBS localization length in the single NW setup we find $\xi'\sim330$~nm. In contrast, the double NW setup with a strength of crossed-Andreev pairing  $\Delta_{c}=0.08$~meV yields a reduction of the MBS localization length to $\xi\sim160$~nm. Inversely, a localization length of $\xi\sim330$~nm which is comparable to the single NW case is achieved already for a Zeeman splitting of $\Delta_{Z}=0.27$~meV corresponding to a field strength of $\sim1$~T. The double NW setup thus allows for MBS localization lengths that are comparable to the single NW setup despite a significant reduction of the magnetic field strength by $\sim1.2$~T.

 \begin{figure}[!t] \centering
\includegraphics[width=\linewidth] {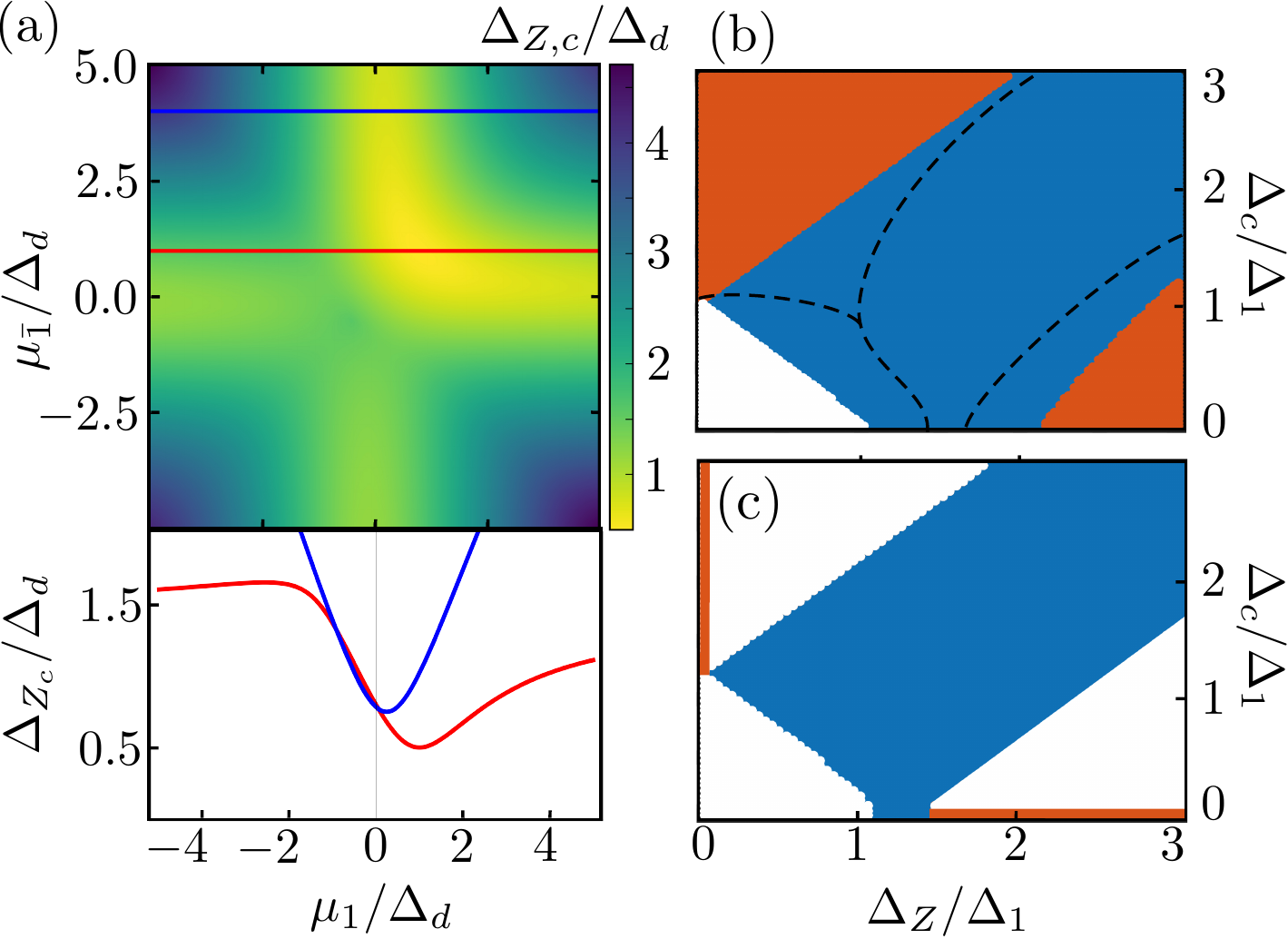}
\caption{(Color online)
(a) Top panel: Color-scale plot of the topological threshold $\Delta_{Z,c}/\Delta_{d}$ for the one-MBS phase as a function of $\mu_{\tau}/\Delta_{d}$ for $\Delta_{c}/\Delta_{d}=0.5$, $\Gamma/\Delta_{d}=1$. Bottom panel: $\Delta_{Z,c}/\Delta_{d}$ as a function of $\mu_{1}/\Delta_{d}$ for $\mu_{\bar{1}}=\Gamma$ (red) and $\mu_{\bar{1}}/\Delta_{d}=4$ (blue). The topological threshold $\Delta_{Z,c}$ exhibits a global minimum for $\mu_{\tau}=\Gamma$.
(b) Topological phase diagram as a function of $\Delta_{Z}/\Delta_{1}$ and $\Delta_{c}/\Delta_{1}$ (obtained from a tight-binding diagonalization of 800 sites per NW) for finite interwire tunneling. We have $E_{so,1}/\Delta_{1}=6.25$,  $E_{so,\bar{1}}/\Delta_{1}=12.25$, $\Delta_{\bar{1}}/\Delta_{1}=1.3$, $\Gamma/\Delta_{1}=1$
, $\mu_{1}=\mu_{\bar{1}}=\Gamma$. Colors are the same as in Fig.~\ref{fig:2}. Black dashed lines denote the approximate phase boundaries for $\mu_{\tau}=0$. For $\mu_{1}=\mu_{\bar{1}}=\Gamma$ the one-MBS phase itself and its phase boundary to the trivial phase remain stable. In contrast, for $\mu_{\tau}=0$, the topological threshold seperating the trivial and one-MBS phase is pushed to higher magnetic fields. 
(c) Same  phase diagram as in (b) but with the two SOI vectors not being parallel to each other but still orthogonal to the magnetic field. We take $\mu_{\tau}=0$, $\Gamma=0$ and $\theta=\pi/6$
for the relative angle between the SOI vectors.
Unlike the one-MBS phase, the two-MBS phases are unstable against rotations of the SOI direction.
}\label{fig:3}
\end{figure}

\section{Stability analysis}
\label{section5}
Next, we study the stability of the one-MBS phase with respect to interwire tunneling and rotations of the SOI vector away from the directions specified in Fig.~\ref{fig:1}. 
First, we show that the effects of interwire tunneling on the low-field topological threshold can be compensated by tuning the NW chemical potentials $\mu_\tau$ to an appropriate sweet spot and we estimate the precision of this tuning. For general  $\mu_\tau$ and finite interwire tunneling, we find that the low-field topological threshold from the trivial to the one-MBS phase occurs at the critical Zeeman splitting 
\begin{align}
\label{Eq8}
\Delta^{2}_{Z,c}&=\Big[2(\Delta_{d}^2+\Delta^{2}_{c}+\Gamma^{2})+\mu^{2}_{1}+\mu^{2}_{\bar{1}}
\\
&\hspace{1pt}- 
\Big(\left[4\Delta_{d}\Delta_{c}\right]^{2}+
\left[4\Delta_{c}^2+(\mu_{1}+\mu_{\bar{1}})^2\right]\left[\mu_{1}-\mu_{\bar{1}}\right]^2\nonumber
\\
&\hspace{20pt}+
4\Gamma\left[\mu_{1}+\mu_{\bar{1}}\right]\left[4\Delta_{d}\Delta_{c}+\Gamma(\mu_{1}+\mu_{\bar{1}})\right]
\Big)^{1/2}\Big]\Big/2.
\nonumber
\end{align}
The critical Zeeman splitting is minimized to $\Delta_{Z,c}=\Delta_{d}-\Delta_{c}$ at the sweet spot $\mu_{\tau}=\Gamma$. For
$|\mu_{\tau}|>\Gamma$, the critical Zeeman splitting increases and approaches $\Delta^{2}_{Z,c}=\Delta^{2}_{d}+\mu^{2}_{\tau}$ when $|\mu_{\bar{\tau}}|\gg|\mu_{\tau}|$, see Fig.~\ref{fig:3}(a) and Fig.~\ref{Fig0_SM} in the Appendix~\ref{AppD}. To tune the chemical potentials to the desired sweet spot, we fix $\mu_{\bar{1}}$ and determine $\Delta_{Z,c}$ as a function of $\mu_{1}$ (e.g. by the emergence of a zero-bias conductance peak). This procedure is repeated for different values of $\mu_{\bar{1}}$. The case $\mu_{\tau}=\Gamma$ is achieved for the global minimum of $\Delta_{Z,c}$ as a function of $\mu_{1}$ and $\mu_{\bar{1}}$.
The required precision of this tuning 
is determined by the width of $\Delta_{Z,c}$ as a function of $\mu_{\tau}$, which is on the scale of $\Delta_{c}$. Importantly, without this tuning the lowering of the topological threshold between the trivial and one-MBS phase does not occur \cite{bib:Gaidamauskas2014} 
as the phase boundary separating the one- and two-MBSs phases shifts to larger magnetic fields, see Fig.~\ref{fig:3}(b).
In Appendix~\ref{AppD} we show that the compensation is still possible in the regime of low Zeeman splittings for $\Delta_{\tau}\sim|\Delta_{1}-\Delta_{\bar{1}}|$ 
but requires an asymmetric tuning of the chemical potentials. 

Second, we address the case when the SOI vectors in the two NWs are not parallel but still orthogonal to the magnetic field vector. 
We replace $H\rightarrow H-i\sum_{\tau,\sigma,\sigma'}\alpha'_{\tau}\int \mathrm{d}x\ \Psi_{\tau\sigma}^\dagger \left(\sigma_y\right)_{\sigma\sigma'}\partial_x\Psi_{\tau\sigma'}$ and set $\alpha_{1}=\alpha$, 
$\alpha'_{1}=0$, $\alpha_{\bar{1}}=\tilde\alpha\cos\theta$, $\alpha'_{\bar{1}}=\tilde\alpha\sin\theta$. 
The new Hamiltonian is in symmetry class D with a $\mathbb{Z}_2$ topological invariant \cite{bib:Ryu2010} and a tight-binding diagonalization reveals a stable one-MBS phase and unstable two-MBSs phases that turn trivial for $\sin\theta\neq0$, see Fig.~\ref{fig:3}(c) and Appendix~\ref{AppE}.

Finally, we have verified by a numerical tight-binding diagonalization as above that the one-MBS phase is stable against Gaussian disorder with mean $\langle\mu_{\tau}\rangle=0$ and a standard deviation that is smaller than the gap. 
\\
\section{Conclusions}
We have shown that in a double NW setup the destructive interference of direct and crossed-Andreev pairing significantly reduces the topological threshold compared to the standard single NW setups \cite{bib:Lutchyn2010,bib:Oreg2010}.
Moreover, we have demonstrated that the resulting MBSs exhibit localization lengths that can be shorter than those of the standard single NW setups. Consequently, they represent ideal candidates for future experiments on quantum information processing with MBSs. 

\section*{Acknowledgments}
We acknowledge support  from the Swiss NSF and NCCR QSIT. We acknowledge helpful discussions with Yanick Volpez.
\\

\appendix
\begin{widetext}
\vspace{0.01in}
\section{
Microscopic model of the proximity effect}
\label{AppA}

In this appendix, we provide a microscopic derivation of the system parameters of our
model Hamiltonian $H=H_{0}+H_{so}+H_{Z}+H_{d}+H_{c}+H_{\Gamma}$ presented in the main text for a weak tunnel coupling between the NWs and the SC, following methods similar to those found in Refs.~\onlinecite{bib:Reeg2017,bib:Reeg2016}. We assume that the NWs are of infinite length, so that the momentum $k_x$ in the $x$ direction is conserved. The momentum-space representation of the bare Hamiltonian of the NWs is given by 
\begin{equation} \label{HNW}
H_0 + H_{so}+H_{Z}=\sum_{\tau,\sigma,\sigma'}\int\frac{dk_x}{2\pi}\,\Psi_{\tau\sigma}^\dagger(k_x)\left(\xi_\tau-\alpha_\tau k_x{\sigma}_z+\Delta_{Z\tau}{\sigma}_x\right)_{\sigma\sigma'}\Psi_{\tau\sigma'}(k_x),
\end{equation}
where we have introduced the single-particle dispersion in the absence of SOI and magnetic field, $\xi_\tau=k_x^2/2m-\mu_\tau$. The Pauli matrices $\sigma_{x,y,z}$ are acting spin space.
The NWs are coupled to an $s$-wave SC of finite width $d$. We describe the SC by the Hamiltonian
\begin{equation} 
\begin{split}
\label{HBCS}
H_{sc}&=\sum_{\sigma}\int\frac{dk_x}{2\pi}\int_0^{d} dz\,
\left[\Psi_{sc,\sigma}^\dagger(k_x,z)\left(-\frac{\partial_z^2}{2m_{sc}}+\frac{k_x^2}{2m_{sc}}-\mu_{sc}\right)\Psi_{sc,\sigma}(k_x,z)\right]
\\&+\frac{\Delta_{sc}}{2}\sum_{\sigma,\sigma'}\int\frac{dk_x}{2\pi}\int_0^{d} dz\,
\left[
 \Psi_{sc,\sigma}^\dagger(k_x,z)
\left(i\sigma_{y}\right)_{\sigma\sigma'}
\Psi_{sc,\sigma'}(-k_x,z)+\text{H.c.}\right]
,
\end{split}
\end{equation}
where $m_{sc}$ and $\mu_{sc}$ are the effective mass and chemical potential of the SC, respectively, and $\Delta_{sc}$ is the superconducting pairing potential. Notably, we neglect the Zeeman splitting due to the applied in-plane field on the superconductor; this is a good approximation at weak magnetic fields if the Zeeman splitting of the superconductor is much smaller than its bulk gap. We also allow for electrons to tunnel locally between SC and NW, assuming that this process preserves both spin and momentum,
\begin{equation}
H_t=-\sum_{\tau,\sigma}\int\frac{dk_x}{2\pi}\,t_\tau\left[\Psi_{\tau\sigma}^\dagger(k_x,z_\tau)\Psi_{sc,\sigma}(k_x,z_\tau)+\text{H.c.}\right],
\end{equation}
where $t_\tau$ is a nanowire-dependent tunneling amplitude and $z_\tau$ denotes the position of the $\tau$-wire. We choose a symmetric configuration $z_1=z_w$ and $z_{\bar{1}}=d-z_w$ while assuming that the wires are located close to the edges of the superconductor, $z_w\ll d$.

The total Hamiltonian can be diagonalized by introducing a Bogoliubov transformation. The resulting Bogoliubov-de Gennes (BdG) equations are given by
\begin{equation}
\begin{split}
&\sum_{\sigma'}\left[\left(\xi_\tau-\alpha_\tau k_x{\sigma}_z+\Delta_{Z\tau}{\sigma}_x\right)_{\sigma\sigma'}u_{\tau\sigma'}(k_x)-t_\tau u_{sc,\sigma}(k_x,z_\tau)\right]=Eu_{\tau\sigma}(k_x), \\
&\sum_{\sigma'}\left[-\left(\xi_\tau-\alpha_\tau k_x{\sigma}_z+\Delta_{Z\tau}{\sigma}_x\right)_{\sigma'\sigma}+t_\tau v_{sc,\sigma}(k_x,z_\tau)\right]=Ev_{\tau\sigma}(k_x), \\
&\sum_{\tau,\sigma'}\left[\left(-\frac{\partial_z^2}{2m_{sc}}+\frac{k_x^2}{2m_{sc}}-\mu_{sc}\right)u_{sc,\sigma}(k_x,z)+i\Delta_{sc}({\sigma}_y)_{\sigma\sigma'}v_{sc,\sigma'}(-k_x,z)-t_\tau\delta(z-z_\tau)u_{\tau\sigma}(k_x)\right]=Eu_{sc,\sigma}(k_x,z), \\
&\sum_{\tau,\sigma'}\left[\left(\frac{\partial_z^2}{2m_{sc}}-\frac{k_x^2}{2m_{sc}}+\mu_{sc}\right)v_{sc,\sigma}(k_x,z)-i\Delta_{sc}({\sigma}_y)_{\sigma\sigma'}u_{sc,\sigma'}(-k_x,z)+t_\tau\delta(z-z_\tau)v_{\tau\sigma}(k_x)\right]=Ev_{sc,\sigma}(k_x,z).
 \label{BdG}
\end{split}
\end{equation}
Here, $u_{\tau[sc]\sigma}(v_{\tau[sc]\sigma})$ is the wave function describing an electron (hole) of spin $\sigma$ in the $\tau$-wire [in the SC]. Inside the SC (\textit{i.e.}, for $0<z<d$), we must solve the BdG equations for a conventional $s$-wave SC:
\begin{equation} \label{BdGS}
\left[\left(-\frac{\partial_z^2}{2m_{sc}}+\frac{k_x^2}{2m_{sc}}-\mu_{sc}\right)\eta_z-\Delta_{sc}\eta_y\sigma_y\right]\psi_{sc}(k_x,z)=E\psi_{sc}(k_x,z),
\end{equation}
where $\psi_{sc}(k_x,z)=[u_{sc,\uparrow}(k_x,z),u_{sc,\downarrow}(k_x,z),v_{sc,\uparrow}(-k_x,z),v_{sc,\downarrow}(-k_x,z)]^T$ is a spinor wave function. Solving independently in the left ($z<z_w$), middle ($z_w<z<d-z_w$), and right ($z>z_w$) regions, the wave function can be expressed as
\begin{align}
\psi_l(z<z_w)&=c_1\chi_{e,\uparrow}\sin(p_+z)+c_2\chi_{e,\downarrow}\sin(p_+z)+c_3\chi_{h,\uparrow}\sin(p_-z)+c_4\chi_{h,\downarrow}\sin(p_-z), \\
\psi_{m}(z_w<z<d-z_w)&=c_5\chi_{e,\uparrow}e^{ip_+z}+c_6\chi_{e,\downarrow}e^{ip_+z}+c_7\chi_{e,\uparrow}e^{-ip_+z}+c_8\chi_{e,\downarrow}e^{-ip_+x}\nonumber \\
&\quad+c_9\chi_{h,\uparrow}e^{ip_-z}+c_{10}\chi_{h,\downarrow}e^{ip_-z}+c_{11}\chi_{h,\uparrow}e^{-ip_-z}+c_{12}\chi_{h,\downarrow}e^{-ip_-z}, \nonumber\\
\psi_r(z>z_w)&=c_{13}\chi_{e,\uparrow}\sin[p_+(d-z)]+c_{14}\chi_{e,\downarrow}\sin[p_+(d-z)]+c_{15}\chi_{h,\uparrow}\sin[p_-(d-z)]+c_{16}\chi_{h,\downarrow}\sin[p_-(d-z)]\nonumber
\end{align}
where $p_\pm^2=2m_{sc}(\mu_{sc}\pm i\Omega)-k_x^2$, with $\Omega^2=\Delta_{sc}^2-E^2$. The spinors are defined as $\chi_{e,\uparrow}=(u_0,0,0,v_0)^T$, $\chi_{e,\downarrow}=(0,u_0,-v_0,0)^T$, $\chi_{h,\uparrow}=(0,-v_0,u_0,0)^T$, and $\chi_{h,\downarrow}=(v_0,0,0,u_0)^T$, where $u_0$ and $v_0$ are the BCS coherence factors,
\begin{equation}
u_0(v_0)=\sqrt{\frac{1}{2}\left(1\pm\frac{i\Omega}{E}\right)}.
\end{equation}
The nanowires enter only through the boundary conditions. These boundary conditions, which must be imposed at $z=z_\tau$ (vanishing boundary conditions at $z=0$ and $z=d$ are accounted for already), are given by
\begin{equation}
\begin{split}
&\psi_l(z_1)=\psi_m(z_1), \\
&\psi_r(z_{\bar{1}})=\psi_m(z_{\bar1}), \\
&[\partial_z\psi_m(z_1)-\partial_x\psi_l(z_1)]/k_{F}=2\gamma_1\eta_zG_1^R(E,k_x)\psi_l(z_1), \\
&[\partial_z\psi_m(z_{\bar1})-\partial_x\psi_r(z_{\bar1})]/k_{F}=2\gamma_{\bar1}\eta_zG_{\bar1}^R(E,k_x)\psi_l(z_{\bar1}),
\label{BC}
\end{split}
\end{equation}
where $G_\tau^R(E,k_x)=(E-\xi_\tau\eta_z+\alpha_\tau k_x\sigma_z-\Delta_{Z\tau}\eta_z\sigma_x+i0^+)^{-1}$ is the retarded Green's function of the $\tau$-wire in the absence of tunneling.
The boundary conditions Eq.~(\ref{BC}) can be rearranged into the form $Mc=0$, where $M$ is a $16\times16$-matrix and $c$ is a 16-component vector of unknown coefficients. The excitation spectrum of the junction is determined by solving the equation $\det{M}=0$ for $E(k_x)$.

We now map the exact BdG solution to the effective pairing model in the limit of weak coupling. We adopt the following approximations. First, we assume that the chemical potential of the SC is the largest energy scale of the problem ($\mu_{sc}\gg E_{so},\Delta_{Z\tau},\Delta_{sc},\mu_\tau$). This allows us to approximate
\begin{equation} \label{approximation}
p_\pm=\sqrt{2m_{sc}(\mu_{sc}\pm i\Omega)-k_x^2}\approx k_{F,sc}\pm i\Omega/v_{F,sc},
\end{equation}
where $k_{F,sc}=\sqrt{2m_{sc}\mu_{sc}}$ and $v_{F,sc}=k_{F,sc}/m_{sc}$ are the Fermi momentum and velocity of the SC, respectively. When differentiating the wave function [on the left-hand side of Eq.~(\ref{BC})], we approximate $p_\pm=k_{F,sc}$; however, in the exponentials [entering through $\psi_{sc}(k_x,z_\tau)$], we keep $p_\pm=k_{F,sc}\pm i\Omega/v_{F,sc}$ (this gives the exponentially growing/decaying parts of the wave function). The weak-tunneling limit is assumed by taking $\gamma_\tau\ll\Delta_{sc}$, where $\gamma_\tau$ is a nanowire-dependent tunneling energy scale given by $\gamma_\tau=t_\tau^2/v_{F,sc}$. In this limit, the relevant pairing energies are small ($E\ll\Delta_{sc}$) and we can expand the coherence factors as
\begin{equation}
u_0(v_0)=\left(\frac{1\pm i}{2}\right)\frac{\Delta_{sc}}{E}.
\end{equation}
However, even with these simplifications, the matrix $M$ [defined below Eq.~(\ref{BC})] is too complicated to be displayed explicitly here. 

Also due to the complicated nature of the matrix $M$, we can only evaluate $\det M$ numerically; this means that the energy spectrum $E(k_x,\mu_\tau,\alpha_\tau,\Delta_{Z\tau},\gamma_\tau,d)$ must be effectively ``guessed" to be mapped out over all of parameter space (i.e., it would be very computationally expensive to numerically map out the spectrum as function of all parameters of the problem). Luckily, it is actually quite straightforward to guess the correct spectrum.

The superconductor induces four effective terms in the Hamiltonian of the NWs. Induced pairing terms are of the direct type,
\begin{equation}
H_d=\sum_{\tau}\Delta_\tau\int\frac{dk_x}{2\pi}\left[\Psi^\dagger_{\tau1}(k_x)\Psi^\dagger_{\tau\bar{1}}(-k_x)+\text{H.c.}\right],
\end{equation}
and the crossed-Andreev type,
\begin{equation}
H_c=\Delta_c\sum_\tau\int\frac{dk_x}{2\pi}\left[\Psi^\dagger_{\tau1}(k_x)\Psi^\dagger_{\bar\tau\bar{1}}(-k_x)+\text{H.c.}\right].
\end{equation}
In addition, the superconductor induces single-particle couplings, which can be of the intrawire type,
\begin{equation}
H_{\delta\mu}=-\sum_{\tau,\sigma}\delta\mu_\tau\int\frac{dk_x}{2\pi}\left[\Psi^\dagger_{\tau\sigma}(k_x)\Psi_{\tau\sigma}(k_x)+\text{H.c.}\right],
\end{equation}
and of the interwire type,
\begin{equation}
H_\Gamma=-\Gamma\sum_{\tau,\sigma}\int\frac{dk_x}{2\pi}\left[\Psi^\dagger_{\tau\sigma}(k_x)\Psi_{\bar\tau\sigma}(k_x)+\text{H.c.}\right].
\end{equation}
With these proximity-induced terms, we propose to describe the nanowires with an effective Hamiltonian of the form
\begin{equation}
H=\frac{1}{2}\int\frac{dk_x}{2\pi}\Phi^\dagger(k_x){\mathcal{H}}_\text{eff}(k_x)\Phi(k_x),
\end{equation}
where ${\Phi}=(\Psi_{11},\Psi_{1\bar{1}},\Psi^\dagger_{11},\Psi^\dagger_{1\bar{1}},\Psi_{\bar11},\Psi_{\bar1\bar{1}},\Psi^\dagger_{\bar11},\Psi^\dagger_{\bar1\bar{1}})^T$ and the effective Hamiltonian density $\hat{\mathcal{H}}_\text{eff}$ is given by
\begin{equation}
\begin{split}
\mathcal{H}(k_{x})
&=
({\xi}_{1}-\delta\mu_1)
\left(\frac{1+\tau_{z}}{2}\right)\eta_{z}
+
({\xi}_{\bar{1}}-\delta\mu_{\bar1})
\left(\frac{1-\tau_{z}}{2}\right)\eta_{z}
-
\alpha_{1}k_{x}
\left(\frac{1+\tau_{z}}{2}\right)\sigma_{z}
-
\alpha_{\bar{1}}k_{x}
\left(\frac{1-\tau_{z}}{2}\right)\sigma_{z}
\\
&\quad+
\Delta_{Z1}
\left(\frac{1+\tau_{z}}{2}\right)\eta_{z}\sigma_{x}
+
\Delta_{Z\bar{1}}
\left(\frac{1-\tau_{z}}{2}\right)\eta_{z}\sigma_{x}
-
\Delta_{c}
\tau_{x}\eta_{y}\sigma_{y}
-
\Gamma
\tau_{x}\eta_{z}
\\
&
\quad-
\Delta_{1}
\left(\frac{1+\tau_{z}}{2}\right)
\eta_{y}\sigma_{y}
-
\Delta_{\bar{1}}
\left(\frac{1-\tau_{z}}{2}\right)
\eta_{y}\sigma_{y}
\label{Heff}
\end{split}
\end{equation}
with the Pauli matrices $\tau_{x,y,z}$, $\eta_{x,y,z}$, $\sigma_{x,y,z}$
acting in nanowire, particle-hole and spin-space, respectively. In the special case when $\Delta_{Z\tau}=0$, the Hamiltonian obeys both time-reversal and particle-hole symmetry with operators $U_{T}=\sigma_{y}$, $U_{P}=\eta_{x}$, and transformations $U^{\dagger}_{T}\mathcal{H}^{*}(k_{x})U_{T}=\mathcal{H}(-k_{x})$, $U^{\dagger}_{P}\mathcal{H}^{*}(k_{x})U_{P}=-\mathcal{H}(-k_{x})$, respectively. Furthermore, $U^{*}_{T}U_{T}=-1$, $U^{*}_{P}U_{P}=1$. Hence, for $\Delta_{Z\tau}=0$ the Hamiltonian is placed in the DIII symmetry class with a $\mathbb{Z}_{2}$ topological invariant \cite{bib:Ryu2010}. 
In general, the Hamiltonian also exhibits an effective time-reversal symmetry described by $U'_{T}=\eta_{z}\sigma_{x}$ with $(U'_{T})^{2}=1$. Therefore, for $\Delta_{Z\tau}\neq0$ the Hamiltonian is placed in the symmetry class BDI with a $\mathbb{Z}$ topological invariant \cite{bib:Ryu2010}. However, we note that the effective time-reversal symmetry $U'_{T}$ is unstable 
when the SOI vectors are not parallel but still orthogonal to the magnetic field vector, 
\begin{equation}
k_{x}\alpha
\left(
\frac{
1+\tau_{z}
}
{2}\right)
\sigma_{z}
+
k_{x}\tilde\alpha\left(
\frac{
1-\tau_{z}
}
{2}\right)
\left[
\cos(\theta)
\sigma_{z}
+
\sin(\theta)
\eta_{z}\sigma_{y}
\right],
\end{equation}
where $\theta\in[0,2\pi)$ is the relative angle between the SOI vectors. Moreover, the effective time-reversal symmetry $U'_{T}$ is also unstable if we allow for a magnetic field vector component that is aligned with one of the SOI vectors, 
\begin{equation}
\Delta_{Z1}\left(
\frac{
1+\tau_{z}
}
{2}\right)
\eta_{z}\sigma_{x}
+
\Delta_{Z\bar{1}}\left(
\frac{
1-\tau_{z}
}
{2}\right)
\eta_{z}
\left[
\cos(\phi)
\sigma_{x}
+
\sin(\phi)
\sigma_{z}
\right].
\end{equation}
with $\phi\in[0,2\pi)$. In the presence of either one of these perturbations with $\sin(\theta)\neq0$ or $\sin(\phi)\neq0$, the Hamiltonian is in the symmetry class D with a $\mathbb{Z}_2$ topological invariant \cite{bib:Ryu2010}. 
\\

The effective parameters $\Delta_\tau$, $\Delta_c$, $\delta\mu_\tau$, and $\Gamma$ were determined as functions of the tunneling strength $\gamma_\tau$ and wire separation $d$ in the absence of spin-orbit coupling and Zeeman splitting in a previous work\cite{bib:Reeg2017}. In the simplified limit $\sin^2(k_Fz_w)=1$, they are given by ($\xi_{sc}=v_{F,sc}/\Delta_{sc}$ is the superconducting coherence length)
\begin{equation}
\begin{split}
&\Delta_\tau=\frac{2\gamma_\tau\sinh(2d/\xi_{sc})}{\cosh(2d/\xi_s)-\cos(2k_Fd)}, \\
&\Delta_c=\frac{4\sqrt{\gamma_1\gamma_{\bar1}}\sinh(d/\xi_{sc})\cos(k_Fd)}{\cosh(2d/\xi_{sc})-\cos(2k_Fd)}, \\
&\delta\mu_\tau=-\frac{2\gamma_\tau\sin(2k_Fd)}{\cosh(2d/\xi_{sc})-\cos(2k_Fd)}, \\
&\Gamma=-\frac{4\sqrt{\gamma_1\gamma_{\bar1}}\cosh(d/\xi_{sc})\sin(k_Fd)}{\cosh(2d/\xi_{sc})-\cos(2k_Fd)}.
\label{effectiveparams}
\end{split}
\end{equation}
Because the effective proximity-induced parameters should depend only on properties of the superconductor and the tunneling amplitude, let us make the ansatz that all four of the proximity-induced effective parameters given in Eq.~(\ref{effectiveparams}) remain unchanged when spin-orbit coupling and a Zeeman splitting are added to the nanowires. That is, we substitute Eq.~(\ref{effectiveparams}) to describe the effective parameters of Eq.~(\ref{Heff}). We then find that if we substitute the energy eigenvalues $E$ of Eq.~(\ref{Heff}) into the boundary conditions Eq.~(\ref{BC}), these choices of $E$ ensure that $\det M=0$; this means that the eigenvalues of the effective Hamiltonian (\ref{Heff}) correspond to the energy spectrum obtained by solving the BdG equations.

\section{Energy spectrum in the strong SOI regime}
\label{AppB}
In this appendix, we compute the bulk energy spectrum of the model $H=H_{0}+H_{so}+H_{Z}+H_{d}+H_{c}$ proposed in the main text \cite{bib:Gaidamauskas2014,bib:Klinovaja2014,Tanaka}. Additionally, we will determine the gapless points of the spectrum which potentially correspond to topological phase boundaries. We assume the regime of strong SOI, $E_{so,\tau}\gg\Delta_{Z},\Delta_{\tau},\Delta_{c}$, and that the deviations in the proximity-induced gaps are the smallest energy scale \cite{bib:Klinovaja2014,Tanaka}, $\Delta_{\tau}\gg|\Delta_{1}-\Delta_{\bar{1}}|$ and $\Delta_{Z\tau}\gg|\Delta_{Z1}-\Delta_{\bar{Z1}}|$. This allows us to set $\Delta_{d}=\Delta_{\tau}$, $\Delta_{Z}=\Delta_{Z\tau}$ and to neglect the effects of interwire tunneling as they can always be compensated by an appropriate adjustment of the nanowire chemical potentials.

We begin by expanding the electron operator according to \cite{Braunecker,peter}
\begin{equation}
\Psi_{\tau\sigma}(x)=R_{\tau\sigma}(x)e^{i\frac{(\sigma+1)}{2}k_{F\tau}x}+L_{\tau\sigma}(x)e^{i\frac{(\sigma-1)}{2}k_{F\tau}x},
\end{equation}
where $R_{\tau\sigma}(x)$ and $L_{\tau\sigma}(x)$ are slowly varying right and left moving fields with spin $\sigma/2$ in the $\tau$-wire. Furthermore, we recall that $k_{F\tau}=2k_{so,\tau}$. Next, we distinguish between three regimes.

\subsection{Strongly detuned SOI energies}
The first case corresponds to strongly detuned nanowire SOI energies, $|E_{so,1}-E_{so,\bar{1}}|\gg\Delta_{Z},\Delta_{d},\Delta_{c}$. In this case the crossed-Andreev pairing couples the two nanowires only at $k=0$. The Hamiltonian 
is given by $H=(1/2)\int \mathrm{d}x \text{ }��\Psi^{\dag}(x)\mathcal{H}(x)\Psi(x)$ with the Hamiltonian density
\begin{align}
\mathcal{H}(x) &=\hbar\upsilon_{F1} \hat k  (1+\tau_z) \rho_z /2 +\hbar\upsilon_{F\bar 1} \hat k (1-\tau_z) \rho_z/2 
+
\Delta_{Z}\eta_{z}(\sigma_{x}\rho_{x}+\sigma_{y}\rho_{y})/2
\\&\quad
+\Delta_{c} \tau_x\eta_y (\sigma_x \rho_y-\sigma_y\rho_x)/2
+\Delta_d \eta_y\sigma_y\rho_x\nonumber 
\end{align}
and the basis $\Psi$=($R_{11}$, $L_{11}$, $R_{1\bar1}$, $L_{1\bar1}$, $R^\dagger_{11}$, $L^\dagger_{11}$, $R^\dagger_{1\bar1}$, $L^\dagger_{1\bar1}$, $R_{\bar 1\bar1}$, $L_{\bar 1 1}$, $R_{\bar1\bar1}$, $L_{\bar1\bar1}$, $R^\dagger_{\bar 1\bar1}$, $L^\dagger_{\bar 1 1}$, $R^\dagger_{\bar1 \bar1}$, $L^\dagger_{\bar 1 \bar1})$. The Pauli matrices $\tau_{x,y,z}$, $\eta_{x,y,z}$, $\sigma_{x,y,z}$, $\rho_{x,y,z}$ act in nanowire, electron-hole, spin and right-left mover space respectively. 
Furthermore, $\hat{k}=-i\partial_{x}$ denotes the momentum operator
whose eigenvalues are $k$ and measured with respect to the
Fermi points at $0,\pm k_{F\tau}=\pm2k_{so,\tau}$ and $\upsilon_{F\tau}=\hbar k_{so,\tau}/m$ is the Fermi velocity in the $\tau$-wire.

We find that the bulk energy spectrum is given by 
\begin{equation}
\begin{split}
&E_{\tau}^2=(\hbar \upsilon_{F\tau} k)^2 + \Delta_d^2,\\
&E_{\pm\pm}^2 = \frac{1}{2} \Bigg[ 
 \hbar^2 \left(\upsilon_{F1}^2+\upsilon_{F\bar{1}}^2\right)k^2
 +2 \left(\Delta_{d}\pm\Delta _Z\right)^2
+2 \Delta _c^2
   \\&\qquad\qquad\pm
\sqrt{
4 \Delta _c^2
 \left(
\hbar ^2 \left[\upsilon _{F1}-\upsilon _{F\bar{1}}\right]^2  k^2
+4 \left[\Delta_{d}\pm\Delta_{Z}\right]^{2}\right)+\hbar^4 \left(\upsilon
   _{F1}^2-\upsilon _{F\bar{1}}^2\right)^2 k^4 }
   \Bigg],
\end{split}
\end{equation}
where the first (second) equation corresponds to exterior (interior) branch of the spectrum.  We find that the spectrum is gapless at $k=0$ provided $\Delta_{c}=|\Delta_{d}\pm\Delta_{Z}|$. There exist no gapless closing points for $k\neq0$.

\subsection{Weakly detuned SOI energies}
The second case corresponds to weakly detuned nanowire SOI energies, $|E_{so,1}-E_{so,\bar{1}}|\ll\Delta_{Z},\Delta_{d},\Delta_{c}$. In this limit, we neglect the difference in spin-orbit energies, $\upsilon_{F}=\upsilon_{F1}=\upsilon_{F\bar{1}}$. The crossed-Andreev pairing now couples the two nanowires both at $k=0$ and $k=\pm k_{F}=\pm2k_{so}$. The Hamiltonian density is given by 
\begin{align}
\mathcal{H}(x) &=\hbar\upsilon_{F} \hat k  \rho_z 
+
\Delta_{Z}\eta_{z}(\sigma_{x}\rho_{x}+\sigma_{y}\rho_{y})/2
-\Delta_{c} \tau_x\eta_y\sigma_y\rho_x
+\Delta_d \eta_y\sigma_y\rho_x
,\nonumber 
\end{align}
and the bulk spectrum is modified to
\begin{equation}
\begin{split}
&E_{\pm}^2= 
( \hbar \upsilon_F k)^2
 +(\Delta_{d}\pm\Delta _c)^2,\\
&E_{\pm\pm}^2 = 
( \hbar \upsilon_{F} k)^2
 +\left(\Delta_c\pm|\Delta_{d}\pm\Delta _Z|\right)^2,
\end{split}
\end{equation}
where the first (second) equation corresponds to exterior (interior) branch of the spectrum.
Besides the gap closing at $k=0$ when $\Delta_{c}=|\Delta_{d}\pm\Delta_{Z}|$, 
we find an additional gap closing at $k=\pm k_{F}=2k_{so}$ when $\Delta_{c}=\Delta_{d}$. For $\Delta_{Z}=0$ this gap closing does not correspond to a topological phase transition because the SOI interaction can be removed by a gauge transformation. For $\Delta_{Z}>0$ we also find from a numerical tight-binding diagonalization that the number of MBS is unchanged
across the gap closing line $\Delta_{c}=\Delta_{d}$, see also Fig.~\ref{Fig5_SM}(b).

\subsection{Intermediate regime}
The last case corresponds to the intermediate regime, when $|E_{so,1}-E_{so,\bar{1}}|\sim\Delta_{Z},\Delta_{d},\Delta_{c}$. To determine the gapless points of the spectrum, we consider for this case the full quadratic Hamiltonian given by $H=(1/2)\int \mathrm{d}x \text{ }��\Phi^{\dag}(x)\mathcal{H}(x)\Phi(x)$ with Hamiltonian density
\begin{equation}
\begin{split}
\mathcal{H}(x)&=
\hbar^{2}\hat{k}^{2}
\eta_{z}/2m
-
\alpha_{1}
\hat{k}
(1+\tau_{z})
\sigma_{z}/2
-
\alpha_{\bar{1}}
\hat{k}
(1-\tau_{z})
\sigma_{z}/2
+
\Delta_{Z}
\eta_{z}\sigma_{x}
-
\Delta_{d}
\eta_{y}\sigma_{y}
-
\Delta_{c}
\tau_{x}\eta_{y}\sigma_{y}
\end{split}
\end{equation}
and basis $\Phi=(\Psi_{11},\Psi_{1\bar{1}},\Psi^{\dag}_{11},\Psi^{\dag}_{1\bar{1}},\Psi_{\bar{1}1},\Psi_{\bar{1}\bar{1}},\Psi^{\dag}_{\bar{1}1},\Psi^{\dag}_{\bar{1}\bar{1}})$. We focus on the gap closing points at finite momentum, because the zero momentum gap closing points are not affected by the SOI. Furthermore, because a finite magnetic field cannot open an energy gap at finite momentum in the regime of strong SOI, we can restrict ourselves to the case when $\Delta_{Z}=0$. Our findings will be equally valid for the case when $\Delta_{Z}\neq0$. First, we determine the bulk energy spectrum,
\begin{equation}
\begin{split}
&
E^{2}_{\pm\pm}(\Delta_{Z}=0)=
\left(
\frac{\hbar^2k^2}{2m}
\right)^{2}
+
\frac{
k^{2}(\alpha^{2}_{1}+\alpha^{2}_{\bar{1}})
}
{2}
+
\Delta^{2}_{d}
+
\Delta^{2}_{c}
\pm
k
(\alpha_{1}+\alpha_{\bar{1}})
\left(
\frac{\hbar^2k^2}{2m}
\right)
\\
&\qquad\qquad\qquad\quad\pm
\sqrt{
\left(
k[\alpha_{1}-\alpha_{\bar{1}}]
\left[
\frac{
k(\alpha_{1}+\alpha_{\bar{1}})
}
{2}
-
\frac{\hbar^2k^2}{2m}
\right]
\right)^{2}
+
\Delta_{c}^2 \left(k^2 [\alpha_{1}-\alpha_{\bar{1}}]^2+4\Delta_{d}^2\right)
}
\end{split}\, \, .
\end{equation}
Next, we find that the spectrum is gapless provided that 
\begin{equation}
\Delta^{2}_{c}
=
\Delta^{2}_{d}
-
\left(
\frac{\hbar^2k^2}{2m}
\right)^{2}
-
k^{2}\alpha_{1}\alpha_{\bar{1}}
+
\left(
\frac{\hbar^2k^2}{2m}
\right)
k
(\alpha_{1}+\alpha_{\bar{1}})
\pm
i\Delta_{d}
\left[
2
\left(
\frac{\hbar^2k^2}{2m}
\right)
-
k
(\alpha_{\bar{1}}
+
\alpha_{1})
\right].
\label{gap_closing_1}
\end{equation}
The spectrum is also gapless for the same condition, but with $k\rightarrow-k$. Because $\Delta_{c}>0$, we need to require that
\begin{equation}
2
\left(
\frac{\hbar^2k^2}{2m}
\right)
-
k
(\alpha_{\bar{1}}
+
\alpha_{1})
=
0.
\end{equation}
Solving this expression (and the corresponding one with $k\rightarrow-k$) for $k$, yields the two gap-closing points
\begin{equation}
k^{*}=\pm
\frac{2m}{\hbar^{2}}
\left(
\frac{\alpha_{\bar{1}}+\alpha_{1}}
{2}
\right).
\end{equation}
Inserting this result back into Eq.~\eqref{gap_closing_1}, we find the gap-closing condition for $k\neq0$,
\begin{equation}
\Delta_{c}=\Delta^{*}_{c}\equiv
\Delta_d\sqrt{
1+4
\left(
\frac{E_{so,1}-E_{so,\bar{1}}}{\Delta_{d}}
\right)
^{2}
}.
\label{finite_k_gap_closing}
\end{equation}
We note that $\Delta^{*}_{c}\geq\Delta_{d}$, so that the gap closing occurs for larger values of the strength of the crossed-Andreev pairing as compared to the regime when $|E_{so,1}-E_{so,\bar{1}}|\ll\Delta_{Z},\Delta_{d},\Delta_{c}$.
Additionally, we emphasize once more that the result in Eq.~\eqref{finite_k_gap_closing} is valid also for $\Delta_{Z}\neq0$ in the limit of strong SOI. Finally, we point out that the topological phase diagram for the regime $|E_{so,1}-E_{so,\bar{1}}|\sim\Delta_{Z},\Delta_{d},\Delta_{c}$ is given in Fig.~\ref{Fig4_SM}(b).

\section{Majorana bound state wavefunctions in the strong SOI regime}
\label{AppC}
In this appendix, we compute the zero-energy MBS wavefunctions of the model $H=H_{0}+H_{so}+H_{Z}+H_{d}+H_{c}$. As in the last section and the main text, we assume the regime of strong SOI, $E_{so,\tau}\gg\Delta_{Z},\Delta_{\tau},\Delta_{c}$, and that the fluctuations in the proximity-induced gaps are the smallest energy scale, $\Delta_{\tau}\gg|\Delta_{1}-\Delta_{\bar{1}}|$ and $\Delta_{Z\tau}\gg|\Delta_{Z1}-\Delta_{\bar{Z1}}|$. This allows us to once again set $\Delta_{d}=\Delta_{\tau}$, $\Delta_{Z}=\Delta_{Z\tau}$ and to neglect the effects of interwire tunneling as they can always be compensated by an appropriate adjustment of the nanowire chemical potentials. 

We begin by assuming that the nanowire length is much longer than the localization length of the MBSs. This means that the MBSs at opposite ends of the system do not overlap and can hence be treated independently. Next, we choose the origin of our coordinate system so that one of the boundaries of the system is located at $x=0$ and focus on this boundary when computing the wavefunctions. We discuss two regimes:  

\subsection{Strongly detuned SOI energies}
The first regime corresponds to strongly detuned nanowire SOI energies, $|E_{so,1}-E_{so,\bar{1}}|\gg\Delta_{Z},\Delta_{d},\Delta_{c}$. Without loss of generality, we choose $\alpha_{\bar{1}}>\alpha_{1}$. 
For $\Delta_{d}+\Delta_{Z}>\Delta_{c}>|\Delta_{d}-\Delta_{Z}|$, we find a single MBS given by $\gamma = \sum_{\tau}\int\mathrm{d}x \ \phi_{\tau}(x)\cdot\Phi_{\tau}(x)$ where $\Phi^{T}_{\tau}=(\Psi_{\tau1}, \Psi_{\tau\bar{1}}, \Psi^{\dag}_{\tau1}, \Psi^{\dag}_{\tau\bar{1}})$ is
the electron spinor and $\phi_{\tau}=(\phi_{\tau1}, \phi_{\tau\bar{1}}, \phi^{*}_{\tau1}, \phi^{*}_{\tau\bar{1}})$
the wavefunction vector in the $\tau$-wire.  The latter is (up to normalization) given by
\begin{equation}
\begin{split}
\phi_{\tau\sigma}(x)&=
\left[
\frac{1+\tau}{2}
+
\left(
\frac{1-\tau}{2}\right)
\frac{
\sqrt{
4 \Delta _c^2 \upsilon _{F1} \upsilon _{F\bar{1}}+\left(\Delta_d-\Delta_Z\right)^2 \left(\upsilon_{F1}-\upsilon_{F\bar{1}}\right)^2
   }
   +\left(\Delta _d-\Delta _Z\right) \left(\upsilon _{F\bar{1}}-\upsilon_{F1}\right)}{2 \Delta _c \upsilon _{F\bar{1}}
 }
   \right]
   \\
   &\quad\times
ie^{i\pi(1-\sigma)/4}
\left(e^{-x/\xi_{e\tau}-i\sigma k_{F\tau} x}-e^{-x/\xi_{i}}\right)
,
\end{split}
\end{equation}
with the localization lengths corresponding to the interior ($i$) and exterior ($e$) branches of the spectrum given by
\begin{equation}
\begin{split}
\xi_{i}&=
\frac{
2 \hbar  \upsilon_{F1} \upsilon_{F\bar{1}}
}
{
\sqrt{
4 \Delta _c^2 \upsilon _{F1} \upsilon_{F\bar{1}}+\left(\Delta _d-\Delta _Z\right)^2 \left(\upsilon _{F1}-\upsilon _{F\bar{1}}\right)^2}+
\left(\Delta_Z-\Delta_d\right) \left(\upsilon_{F1}+\upsilon_{F\bar{1}}\right)
   }.
\end{split}
\end{equation}

For $\Delta_{Z}>\Delta_{d}$ and $\Delta_{c}<\Delta_{Z}-\Delta_{d}$, we find two MBSs given by 
$\gamma = \sum_{\tau}\int\mathrm{d}x \ \phi_{\tau}(x)\cdot\Phi_{\tau}(x)$ and $\gamma' = \sum_{\tau}\int\mathrm{d}x \ \phi'_{\tau}(x)\cdot\Phi_{\tau}(x)$ where the wavefunction vector  $\phi'_{\tau}=(\phi'_{\tau1}, \phi'_{\tau\bar{1}}, (\phi'_{\tau1})^{*}, (\phi'_{\tau\bar{1}})^{*})$ is (up to normalization) given by 
\begin{equation}
\begin{split}
\phi'_{\tau\sigma}(x)&=
\left[
\frac{\tau-1}{2}
+
\left(
\frac{1+\tau}{2}\right)
\frac{\sqrt{4 \Delta _c^2 \upsilon _{F1} \upsilon _{F\bar{1}}+\left(\Delta _d-\Delta _Z\right)^2 \left(\upsilon_{F1}-\upsilon_{F\bar{1}}\right)^2}+\left(\Delta _Z-\Delta _d\right) \left(\upsilon _{F1}-\upsilon_{F\bar{1}}\right)}{2 \Delta _c \upsilon _{F1}}
\right]
\\
&\quad\times
ie^{i\pi(1-\sigma)/4}
\left(e^{-x/\xi_{e\tau}-i\sigma k_{F\tau} x}-e^{-x/\xi'_{i}}\right),
\end{split}
\end{equation}
with the localization length
\begin{equation}
\begin{split}
\xi'_{i}&=
\frac{
2 \hbar  \upsilon _{F1} \upsilon _{F\bar{1}}
}{
\left(\Delta _Z-\Delta _d\right) \left(\upsilon _{F1}+\upsilon_{F\bar{1}}\right)-\sqrt{4 \Delta_c^2 \upsilon _{F1} \upsilon _{F\bar{1}}+\left(\Delta_d-\Delta_Z\right)^2
   \left(\upsilon_{F1}-\upsilon_{F\bar{1}}\right)^2}}.
\end{split}
\end{equation}

For $\Delta_{c}>\Delta_{d}+\Delta_{Z}$, we again find two MBSs. They are given by $\gamma = \sum_{\tau}\int\mathrm{d}x \ \phi_{\tau}(x)\cdot\Phi_{\tau}(x)$ and $\gamma'' = \sum_{\tau}\int\mathrm{d}x \ \phi''_{\tau}(x)\cdot\Phi_{\tau}(x)$, where the wavefunction vector  $\phi''_{\tau}=(\phi''_{\tau1}, \phi''_{\tau\bar{1}}, (\phi''_{\tau1})^{*}, (\phi''_{\tau\bar{1}})^{*})$ is (up to normalization) given by 
\begin{equation}
\begin{split}
\phi''_{\tau\sigma}(x)&=
\left[
\frac{1-\tau}{2}
+
\left(
\frac{1+\tau}{2}\right)
\frac{\sqrt{4 \Delta _c^2 \upsilon _{F1} \upsilon _{F\bar{1}}+\left(\Delta_d+\Delta _Z\right)^2 \left(\upsilon_{F1}-\upsilon_{F\bar{1}}\right)^2}+\left(\Delta_d+\Delta_Z\right) \left(\upsilon_{F1}-\upsilon_{F\bar{1}}\right)}{2 \Delta_c \upsilon_{F1}}
\right]
\\
&\quad\times
e^{i\pi(\sigma-1)/4}
\left(e^{-x/\xi_{e\tau}-i\sigma k_{F\tau} x}-e^{-x/\xi''_{i}}\right),
\end{split}
\end{equation}
with the localization length
\begin{equation}
\begin{split}
\xi''_{i}&=
\frac{2 \hbar  \upsilon _{F1} \upsilon _{F\bar{1}}}{\sqrt{4 \Delta_c^2 \upsilon _{F1} \upsilon_{F\bar{1}}+\left(\Delta_d+\Delta_Z\right)^2 \left(\upsilon_{F1}-\upsilon _{F\bar{1}}\right)^2}-\left(\Delta_d+\Delta_Z\right) \left(\upsilon_{F1}+\upsilon _{F\bar{1}}\right)}
.
\end{split}
\end{equation}
We point out that the found MBSs are orthogonal to each other and correspond to independent solutions of the Hamiltonian, because $\sum_{\tau}\phi_{\tau}(x)\cdot\phi'_{\tau}(x)=0$
and $\sum_{\tau}\phi_{\tau}(x)\cdot\phi''_{\tau}(x)=0$. We also note that the remaining parameter regimes which we did not discuss here correspond to topologically trivial phases.

\subsection{Weakly detuned SOI energies}
The second regime corresponds to weakly detuned nanowire SOI energies, $|E_{so,1}-E_{so,\bar{1}}|\ll\Delta_{Z},\Delta_{d},\Delta_{c}$. For simplicity, we assume that $E_{so,1}=E_{so,\bar{1}}$. For $\Delta_{d}+\Delta_{Z}>\Delta_{c}>|\Delta_{d}-\Delta_{Z}|$ and $\Delta_{c}\neq\Delta_{d}$, we find a single MBS given by $\gamma = \sum_{\tau}\int\mathrm{d}x \ \phi_{\tau}(x)\cdot\Phi_{\tau}(x)$ with the wavefunction vector $\phi_{\tau}=(\phi_{\tau1}, \phi_{\tau\bar{1}}, \phi^{*}_{\tau1}, \phi^{*}_{\tau\bar{1}})$
in the $\tau$-wire given (up to normalization) by
\begin{equation}
\begin{split}
\phi_{\tau\sigma}(x)&=
ie^{i\pi(1-\sigma)/4}
\left(e^{-x/\xi_{e}-i\sigma k_{F} x}-e^{-x/\xi_{i}}\right)
\end{split}
\end{equation}
and the localization lengths corresponding to the interior ($i$) and exterior ($e$) branches of the spectrum
\begin{equation}
\begin{split}
\xi_{i}=
\begin{cases} 
\frac{
\hbar\upsilon_{F}
}
{
\Delta_{Z}-(\Delta_{c}-\Delta_{d})
}&\mbox{if } \Delta_{c}>\Delta_{d}  \\ 
\frac{
\hbar\upsilon_{F}
}
{
\Delta_{Z}-(\Delta_{d}-\Delta_{c})
}
& \mbox{if } \Delta_{c}<\Delta_{d} \end{cases}
, \quad
\xi_{e}=\begin{cases} 
\frac{
\hbar\upsilon_{F}
}
{
\Delta_{c}-\Delta_{d}
}&\mbox{if } \Delta_{c}>\Delta_{d}  \\ 
\frac{
\hbar\upsilon_{F}
}
{
\Delta_{d}-\Delta_{c}
}
& \mbox{if } \Delta_{c}<\Delta_{d}. \end{cases}
\end{split}
\end{equation}
For $\Delta_{c}<|\Delta_{d}-\Delta_{Z}|$, $\Delta_{Z}>\Delta_{d}$ and $\Delta_{c}\neq\Delta_{d}$ we find two MBSs 
given by $\gamma = \sum_{\tau}\int\mathrm{d}x \ \phi_{\tau}(x)\cdot\Phi_{\tau}(x)$ and $\gamma' = \sum_{\tau}\int\mathrm{d}x \ \phi'_{\tau}(x)\cdot\Phi_{\tau}(x)$ where the wavefunction vector  $\phi'_{\tau}=(\phi'_{\tau1}, \phi'_{\tau\bar{1}}, (\phi'_{\tau1})^{*}, (\phi'_{\tau\bar{1}})^{*})$ is (up to normalization) given by 
\begin{equation}
\begin{split}
\phi'_{\tau\sigma}(x)&=
i\tau e^{i\pi(1-\sigma)/4}
\left(e^{-x/\xi'_{e}-i\sigma k_{F} x}-e^{-x/\xi'_{i}}\right),
\end{split}
\end{equation}
with the localization lengths 
\begin{equation}
\begin{split}
\xi'_{i}&=\frac{
\hbar\upsilon_{F}
}
{
\Delta_{Z}-(\Delta_{c}+\Delta_{d})
}, \qquad \qquad \quad
\xi'_{e}=\frac{
2\hbar\upsilon_{F}
}
{
\Delta_{c}+\Delta_{d}
}.
\end{split}
\end{equation}
We point out that the solutions for the two-MBS phase are independent, because $\phi_{\tau}(x)\cdot\phi'_{\tau}(x)=0$. The parameter regimes which were not discussed correspond to topologically trivial phases.

\section{Interwire tunneling}
\label{AppD}

In this appendix, we study the effects of tunneling between NWs in the model which we presented in the main text. These interwire tunneling processes are described by 
\begin{equation}
H_{\Gamma} = -\Gamma\sum_{\tau,\sigma} \int \mathrm{d}x \left[ \Psi^\dagger_{\tau\sigma}(x)\Psi_{\bar \tau\sigma}(x)+\text{H.c.}\right],
\end{equation}
where $\Gamma>0$ is a spin-independent tunneling amplitude.
The full Hamiltonian of our system in then given by $H=H_{0}+H_{so}+H_{Z}+H_{d}+H_{c}+H_{\Gamma}$. In this section, we will analytically show that: (1) The effects of interwire tunneling on the topological phase transition between the trivial and the one-MBS phase can always be compensated by an appropriate adjustment of the nanowire chemical potentials when 
$\Delta_{Z\tau},\Delta_{\tau}\gg|\Delta_{1}-\Delta_{\bar{1}}|,|\Delta_{Z1}-\Delta_{Z\bar{1}}|$. For low Zeeman splittings, 
$\Delta_{Z\tau}\ll\Delta_{\tau}$, this compensation is possible even if $\Delta_{Z\tau}\sim|\Delta_{Z1}-\Delta_{Z\bar{1}}|$ and $\Delta_{\tau}\sim|\Delta_{1}-\Delta_{\bar{1}}|$. (2) The latter adjustment of the nanowire chemical potentials expands the one-MBS phase by pushing the topological threshold from the one-MBS into the two-MBS phase to higher Zeeman splittings. 
\\

We first discuss the limit when $\Delta_{Z\tau},\Delta_{\tau}\gg|\Delta_{1}-\Delta_{\bar{1}}|,|\Delta_{Z1}-\Delta_{Z\bar{1}}|$. 
As a starting point, we set $\Delta_{Z}=\Delta_{Z\tau}$, $\Delta_{d}=\Delta_{\tau}$ and redisplay the full Hamiltonian in the presence of interwire tunneling,
$H=\int \mathrm{d}x \text{ }��\Psi^{\dag}(x)\mathcal{H}(x)\Psi(x)$/2 with $\Psi^{\dag}=(\Psi^{\dag}_{11},\Psi^{\dag}_{1\bar{1}},\Psi_{11},\Psi_{1\bar{1}},\Psi^{\dag}_{\bar{1}1},\Psi^{\dag}_{\bar{1}\bar{1}},\Psi_{\bar{1}1},\Psi_{\bar{1}\bar{1}})$ and the Hamiltonian density
\begin{equation}
\begin{split}
\mathcal{H}(x)&=
\left(
\frac{\hbar^2\hat{k}^{2}}{2m}-\mu
\right)
\eta_{z}
-
\alpha
\hat{k}
\sigma_{z}
+
\Delta_{Z}
\eta_{z}\sigma_{x}
-
\Delta_{d}
\eta_{y}\sigma_{y}
-
\Delta_{c}
\tau_{x}\eta_{y}\sigma_{y}
-
\Gamma
\tau_{x}\eta_{z}
.
\end{split}
\end{equation}
The Pauli matrices $\sigma_{x,y,z}$, $\tau_{x,y,z}$, and $\eta_{x,y,z}$ act in spin, nanowire and electron-hole space, respectively. Because we are solely interested in the modification of the zero-momentum gap closing condition $\Delta^{2}_{c}=(\Delta_{d}\pm\Delta_{Z})$ for finite interwire tunneling, we have also set $\alpha=\alpha_{1}=\alpha_{\bar{1}}$. Our model can now be mapped onto a model of effectively two
decoupled topological NWs. To see this, we introduce the basis
\begin{equation}
\widetilde\Psi^{\dag}_{\lambda}=
(
\Psi^{\dag}_{\bar{1}1}+\lambda\Psi^{\dag}_{11}, 
\Psi^{\dag}_{\bar{1}\bar{1}}+\lambda\Psi^{\dag}_{1\bar{1}},
\Psi_{\bar{1}1}+\lambda\Psi_{11},
\Psi_{\bar{1}\bar{1}}+\lambda\Psi_{1\bar{1}}
)/\sqrt{2}.
\end{equation}
We will interpret $\lambda=\pm1$ as an effective nanowire index that  (together with the spin index) labels the energy bands of the system in the absence of superconductivity and magnetic fields fields, $\Delta_{d}=\Delta_{c}=\Delta_{Z}=0$. We choose $\mu=\Gamma$. In this new basis the Hamiltonian density can then be rewritten as $H=-\sum_{\lambda}\int \mathrm{d}x \text{ }��\widetilde\Psi_{\lambda}^{\dag}(x)\widetilde{\mathcal{H}}_{\lambda}(x)\widetilde\Psi_{\lambda}(x)/2$ with
\begin{equation}
\begin{split}
\widetilde{\mathcal{H}}_{\lambda}(x)&=
\left(
\frac{\hbar^2\hat{k}^{2}}{2m}-\mu_{\text{eff},\lambda}
\right)
\eta_{z}
-
\lambda
\alpha
\hat{k}
\sigma_{z}
+
\Delta_{Z}
\eta_{z}\sigma_{x}
-
\lambda
\Delta_{\text{eff},\lambda}
\eta_{y}\sigma_{y}
\end{split}
\end{equation}
where we have introduced the effective chemical potentials $\mu_{\text{eff},1}=0$ and $\mu_{\text{eff},\bar{1}}=2\Gamma$ as well as the effective pairing potentials $\Delta_{\text{eff},1}=\Delta_{d}-\Delta_{c}$ and $\Delta_{\text{eff},\bar{1}}=\Delta_{d}+\Delta_{c}$. This is precisely the Hamiltonian of two decoupled topological NWs labeled by the effective nanowire index $\lambda$.
Thus, the system hosts  one MBS at each end for low magnetic fields when 
\begin{equation}
\Delta^{2}_{Z}>\Delta^{2}_{\text{eff,}1}+\mu^{2}_{\text{eff,}1}=(\Delta_{d}-\Delta_{c})^{2}
\end{equation}
and two MBSs at each end for large magnetic fields when
\begin{equation}
\Delta^{2}_{Z}>\Delta^{2}_{\text{eff,}\bar{1}}+\mu^{2}_{\text{eff,}\bar{1}}=
(\Delta_{d}+\Delta_{c})^{2}
+
(2\Gamma)^{2}.
\end{equation}
Consequently, by an appropriate adjustment of the nanowire chemical potentials (for example by an external gate voltage),
we still observe the proposed one-MBS phase for low magnetic fields. Also, the one-MBS phase now even extends to significantly stronger fields.
Moreover, without a proper tuning of the chemical potentials, $|\mu|\gg\Gamma$, the topological threshold is shifted to higher magnetic fields and not much is gained, see also Fig.~\ref{Fig0_SM}(a) and (b).
\\
\\
Next, we comment on the case when $\Delta_{\tau}\sim|\Delta_{1}-\Delta_{\bar{1}}|$. In this case the choice 
\begin{equation}
\mu_{1} 
=
\Gamma
\sqrt{
\frac{\Delta_{1}}{\Delta_{\bar{1}}}
}
\quad
\text{and}
\quad
\mu_{\bar{1}} 
=
\Gamma
\sqrt{
\frac{\Delta_{\bar{1}}}{\Delta_{1}}
}
\end{equation}
\begin{figure}[!htb]
\includegraphics[width=0.95\columnwidth]{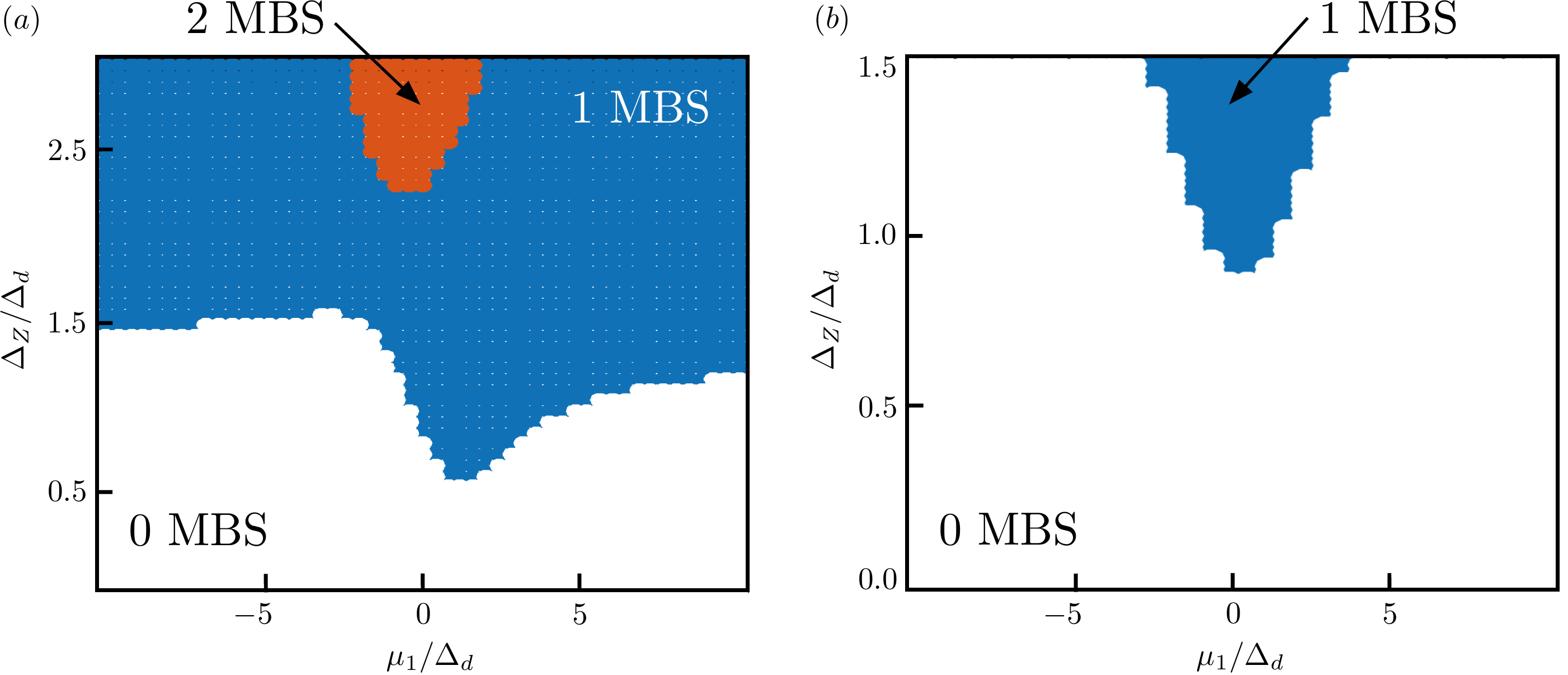}
\caption{
(a) Topological phase diagram as a function of $\Delta_{Z}/\Delta_{d}$ and $\mu_{1}/\Delta_{d}$ (obtained from a tight-binding diagonalization of 800 sites per NW) for finite interwire tunneling, $\Gamma/\Delta_{d}=1$. We have set
$E_{so,\tau}/\Delta_{d}=6.25$, $\Delta_{c}/\Delta_{d}=0.5$ and $\mu_{\bar{1}}/\Delta_{d}=1$
. Colors are the same as in Fig.~2 of the main text. We see that without tuning the chemical potentials to the sweet spot $\mu_{\tau}=\Gamma$ the topological threshold $\Delta_{Z,c}$ is shifted to significantly higher magnetic fields and not much is gained in a double NW setup compared to single NWs. 
(b) Same topological phase diagram as in (a) but for $\mu_{\bar{1}}/\Delta_{d}=4$. Once more, we see that without the tuning to the sweet spot $\mu_{\tau}=\Gamma$ no low-field topological threshold is observed.
 }
\label{Fig0_SM}
\end{figure}
ensures that the effects of interwire tunneling can still be compensated provided $\Delta_{Z\tau}\ll\Delta_{\tau}$. 
However, we note that the topological phase transition from the trivial to the one-MBS phase at $\Delta_{c}=0$ is renormalized to 
\begin{equation}
\label{modified_gap_closing}
\Delta^{2}_{Z,c}=
\frac{1}{2} 
\left[
\Delta_{1}^2
+
\Delta_{\bar{1}}^2
+
\Gamma^{2}
\left(
2
+
\frac{\Delta_{1}}{\Delta_{\bar{1}}}
+
\frac{\Delta_{\bar{1}}}{\Delta_{1}}
\right)
-\frac{
\sqrt{\left(\Delta_{1}\Delta_{\bar{1}} \left[\Delta_{1}^2+\Delta_{\bar{1}}^2\right]+\Gamma^2 \left[\Delta_{1}+\Delta_{\bar{1}}\right]^2\right)^2-4 \Delta_{1}^3 \Delta_{\bar{1}}^3 \left(\Delta_{1}\Delta_{\bar{1}}+4
   \Gamma^2\right)}
   }{
   \Delta_{1}\Delta_{\bar{1}}
   }\right]\geq0.
\end{equation}

\section{Numerical results}
\label{AppE}

In this final section, we study the tight-binding model which corresponds to the continuum model presented in the main text \cite{bib:Prada2012,diego,soi}. 
The tight-binding Hamiltonian is given by 
\begin{equation}
\begin{split}
H&=\left[\sum_{\tau} \left( \sum_{j=1}^{N} \tilde{\psi}_{\tau,j}^\dagger[-\mu_\tau \eta_z + \Delta_\tau \eta_x + \Delta_{Z\tau} \sigma_x]\tilde{\psi}_{\tau,j} +\sum_{j=1}^{N-1} \tilde{\psi}_{\tau,j+1}^\dagger[-t-i\alpha_{\tau} \sigma_z ]~ \eta_z \tilde{\psi}_{\tau,j}+\text{H.c.}\right)\right]
\\
&\quad+
\sum_{j=1}^N\tilde{\psi}_{\bar{1},j}^\dagger~ (\Delta_c \eta_x)~ \tilde{\psi}_{1,j} +\text{H.c.}
,
\label{H0}
\end{split}
\end{equation}
where $N$ is the number of lattice sites per wire and $\tilde{\psi}_{\tau,j}= (\psi_{\tau,j,\uparrow}^\dagger, \psi_{\tau,j,\downarrow}^\dagger, \psi_{\tau,j,\downarrow}, -\psi_{\tau,j,\uparrow})$ is the electron spinor at site $j$ in the $\tau$-wire with
$\psi_{\tau,j,\sigma}$ the annihilation operator of a spin $\sigma$ electron at site $j$ in the $\tau$-wire. Moreover, $\mu_\tau$, $\Delta_\tau$, $\alpha_{\tau}$, $\Delta_{Z\tau}$  are the chemical potentials, direct pairing strengths, SOI strengths and Zeeman splittings in wire $\tau$, respectively. 
Finally, $\Delta_{c}$ is the strength of the crossed-Andreev pairing, $t$ is the hopping amplitude and the Pauli matrices $\sigma_{x,y,z}$ ($\eta_{x,y,z}$) act in spin (particle-hole) space.

\begin{figure}[t]
\includegraphics[width=0.95\columnwidth]{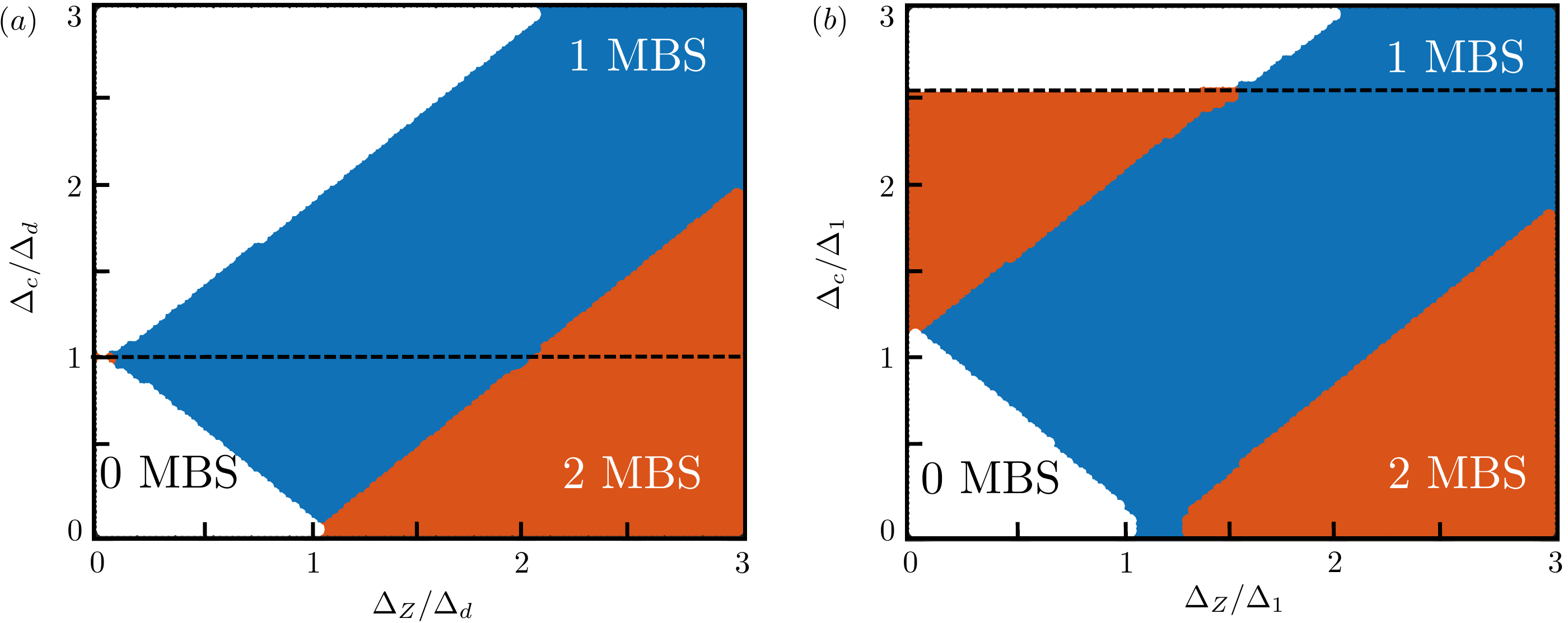}
\caption{(a) Topological phase diagram as a function of $\Delta_{Z}/\Delta_{d}$ and $\Delta_{c}/\Delta_{d}$ for the regime of weakly detuned SOI energies, $|E_{so,1}-E_{so,\bar{1}}| \ll \Delta_{Z\tau},\Delta_{\tau},\Delta_{c}$. The color coding scheme is the same as in Fig.~2 in the main text. The dashed black line denotes a gap closing at finite momentum. We have chosen $E_{so,\tau}/\Delta_{d}=2$ and
 $\mu_1=\mu_{\bar{1}}=0$. 
The two-MBS phase which appeared for $\Delta_{c}>\Delta_{d}+\Delta_{Z}$ when $|E_{so,1}-E_{so,\bar{1}}| \gg \Delta_{Z\tau},\Delta_{\tau},\Delta_{c}$ turns into a trivial phase. All other topological phases remain unchanged.
 (b) Topological phase diagram as a function of $\Delta_{Z}/\Delta_{1}$ and $\Delta_{c}/\Delta_{1}$ for the regime,  $|E_{so,1}-E_{so,\bar{1}}| \sim \Delta_{Z\tau},\Delta_{\tau},\Delta_{c}$. We have chosen $E_{so,1}/\Delta_{1}=2$ ,$E_{so,\bar 1}/\Delta_{1}=4.5$,
 $\mu_1=\mu_{\bar{1}}=0$, $\Delta_{\bar{1}}/\Delta_{1}=1.3$.
  We note that both two-MBS phases disappear for $\Delta_{c}>\Delta^{*}_{c}$, see  Eq.~\eqref{finite_k_gap_closing}.
 }
\label{Fig4_SM}
\end{figure}

\subsection{Topological phase diagram}  
First, we perform a numerical diagonalization to obtain the topological phase diagram for the regime
of weakly detuned SOI energies, $|E_{so,1}-E_{so,\bar{1}}| \ll \Delta_{Z\tau},\Delta_{\tau},\Delta_{c}$, and for the intermediate regime, $|E_{so,1}-E_{so,\bar{1}}| \sim \Delta_{Z\tau},\Delta_{\tau},\Delta_{c}$. The results are shown
in Fig.~\ref{Fig4_SM}. In the limit of weakly detuned SOI energies, we find that the
two-MBS phase which for $|E_{so,1}-E_{so,\bar{1}}| \gg \Delta_{Z\tau},\Delta_{\tau},\Delta_{c}$ with $\Delta_{d}=\Delta_{\tau}$, $\Delta_{Z}=\Delta_{Z\tau}$ appeared when $\Delta_{c}>\Delta_{d}+\Delta_{Z}$, completely turns into a topologically trivial phase, see Fig.~\ref{Fig4_SM}(a). Compared to that, in the intermediate regime, we find that the same two-MBS phase turns into a trivial phase once $\Delta_{c}>\Delta^{*}_{c}$ where $\Delta^{*}_{c}$ was defined in Eq.~\eqref{finite_k_gap_closing}, see Fig.~\ref{Fig4_SM}(b).

\subsection{Stability Analysis} 

Second, we analyze the stability of the one-MBS phase against different perturbations. 

\begin{figure}[t]
\includegraphics[width=0.95\columnwidth]{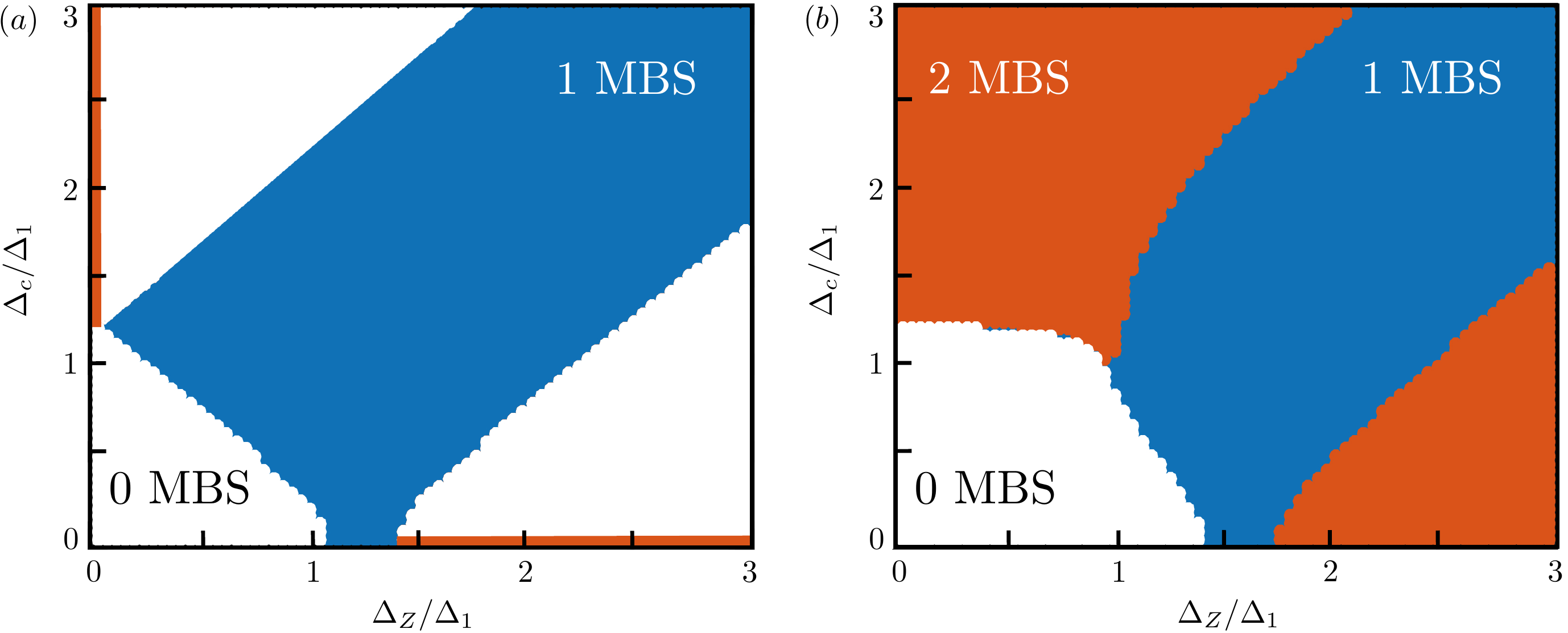}
\caption{(a) Topological phase diagram as a function of $\Delta_{Z}/\Delta_{1}$ and $\Delta_{c}/\Delta_{1}$ for the regime of strongly detuned SOI energies, $|E_{so,1}-E_{so,\bar{1}}| \gg \Delta_{Z\tau},\Delta_{\tau},\Delta_{c}$, and a rotation of the magnetic field in the $\bar{1}$-wire by $\phi=0.2$. The color coding scheme is the same as in Fig.~2 in the main text. We have chosen $E_{so,1}/\Delta_{1}=6.25$, $E_{so,\bar 1}/\Delta_{1}=12.25$, $N=800$,
$\Delta_{\bar{1}}/\Delta_{1}=1.3$, $\mu_1=0 ~\text{and}~ \mu_{\bar{1}}=0$. While the one-MBS remains stable, the two-MBS phases are unstable against rotations of the magnetic field with $\sin(\phi)\neq0$.
(b) Same topological phase diagram as in (a) but with finite interwire tunneling, $\Gamma/\Delta_{1}=1$. Moreover, we also set $\phi=0$ and $\mu_{\tau}=0$. Consequently, the effects of interwire tunneling are [unlike in Fig.~3(b) of the main text] not compensated and the topological threshold from the trivial to the one-MBS phase is pushed to substantially higher magnetic fields. Thus, to get the maximum advantage of the double nanowire setup, it is crucial to compensate for these shifts due to interwire tunneling.
}
\label{Fig5_SM}
\end{figure}

{\it  Misalignments of the magnetic fields}.
First, we discuss rotations of the magnetic field in the $x-z$ plane for the regime of strongly detuned SOI energies, $|E_{so,1}-E_{so,\bar{1}}| \gg \Delta_{Z\tau},\Delta_{\tau},\Delta_{c}$. 
We replace our tight-binding Hamiltonian according to 
\begin{align}
 H \rightarrow H + \sum_{\tau}\sum_{j=1}^N\tilde{\psi}_{\tau,j}^\dagger~ (\Delta_{Z\tau}' \sigma_z)~ \tilde{\psi}_{\tau,j},
\end{align}
and set $\Delta_{Z1}=0, \Delta_{Z1}'=\Delta_Z$ for the $1$-wire and $\Delta_{Z\bar{1}}=\Delta_Z \cos(\phi), \Delta_{Z\bar{1}}'=\Delta_Z \sin(\phi)$ for the $\bar{1}$-wire where $\phi\in[0,2\pi)$ is the angle of the magnetic field acting on the $\bar{1}$-wire with respect to the $x$ axis in the $x-z$ plane. For $\sin(\phi)\neq0$, this places the setup in symmetry class D with a $\mathbb{Z}_2$ topological invariant \cite{bib:Ryu2010}. From a numerical tight-binding diagonalization, we  find that the one-MBS phase remains stable, while the two-MBS phases turn into trivial phases for $\sin(\phi)\neq0$, see Fig.~\ref{Fig5_SM}(a). Additionally, we observe that the one-MBS phase expands to larger magnetic fields. 
\\
\\
{\it Misalignments of the SOI vectors}. 
The case of misaligned SOI vectors in the two wires was discussed in the main text. To obtain the topological phase diagram shown in Fig.~3(c) in the main text, we modify our tight-binding Hamiltonian according to 
\begin{align}
 H \rightarrow H 
+i \sum_{\tau}\sum_{j=1}^N\tilde{\psi}_{\tau,j+1}^\dagger~ \alpha'_{\tau} \sigma_y\eta_z \tilde{\psi}_{\tau,j}, 
\end{align}
and set $\alpha_{1}=\alpha$, $\alpha'_{1}=0$ for the $1$-wire and $\alpha_{\bar{1}}=\tilde\alpha\cos(\theta)$, $\alpha'_{\bar{1}}=\tilde\alpha\sin(\theta)$ for the $\bar{1}$-wire with $\theta$ being the angle of the SOI vector
in the $\bar{1}$-wire relative to the $z$-axis in the $yz$-plane. As a result, we confirm that the one-MBS phase remains stable against misalignments of the SOI vectors. In contrast to that, the two-MBS phase is unstable except special line $\Delta_Z=0$, where time-reversal symmetry guarantees the presence of Kramers doublets \cite{bib:Gaidamauskas2014,bib:Klinovaja2014,Tanaka,Fu,nagaosa,law,nagaosa2,mele,loss,loss2,pikulin}. If $\Delta_Z\neq0$, the two MFs localized at the same end are protected from hybridization by some additional symmetry. However, as noticed above such effective time-reversal symmetries are not stable against general perturbations \cite{pascal,chen,silas,bena}, resulting in lifting of the degeneracy of two zero-energy bound states.
\\
\\
{\it Interwire tunneling}.  Lastly, we provide additional information on our analysis for the case of finite interwire tunneling presented in the main text. In this case, the tight-binding Hamiltonian is modified according to
\begin{align}
H \rightarrow H + \sum_{j=1}^{N} \tilde{\psi}_{2,j}^\dagger (-\Gamma \eta_z) \tilde{\psi}_{1,j}+\text{H.c.},
\end{align}
where $\Gamma>0$ is the spin-independent tunneling amplitude. As discussed in the previous section, we find that the effects of interwire tunneling on the topological phase transition separating the trivial and one-MBS phase can be completely compensated by setting $\mu_{\tau}=\Gamma$. Without this tuning the topological threshold separating the trivial and one-MBS phase is pushed to significantly larger magnetic fields, see see Fig.~\ref{Fig5_SM}(b).

\end{widetext}

\end{document}